\documentclass[aps,twocolumn,showpacs,amsmath,amssymb,prb,footinbib]{revtex4-1}

\usepackage{graphicx}
\usepackage{color}
\usepackage{amsmath}
\usepackage{psfrag}
\usepackage{enumitem}
\usepackage{natbib}
\usepackage{float}

\newcommand{\charge}{c}

\begin{document}

\title{Zero-frequency noise in adiabatically driven, interacting quantum systems}
\author{Roman-Pascal~Riwar}
\author{Janine~Splettstoesser}
\affiliation{Institut f\"{u}r Theorie der Statistischen Physik, RWTH Aachen University, D-52056 Aachen, Germany}
\affiliation{JARA - Fundamentals of Future Information Technology}
\author{J\"urgen~K\"onig}
\affiliation{Theoretische Physik, Universit\"at Duisburg-Essen and CENIDE, D-47048 Duisburg, Germany}
\date{\today}

\begin{abstract}
We investigate current-current correlations of adiabatic charge pumping through interacting quantum dots weakly coupled to reservoirs. 
To calculate the zero-frequency noise for a time-dependently driven system, possibly in the presence of an additional dc bias, we perform within a real-time diagrammatic approach a perturbative expansion in the tunnel coupling to the reservoirs in leading and next-to-leading order.
We apply this formalism to study the adiabatic correction to the zero-frequency noise, i.e., the pumping noise, in
the case of a single-level quantum dot charge pump.
If no stationary bias is applied, the adiabatic correction shows Coulomb-interaction-induced deviations from the fluctuation-dissipation theorem. Furthermore, we show that the adiabatic correction to the Fano factor carries information about the coupling asymmetry and is independent of the choice of the pumping parameters.
When including a time-dependent finite bias, we find that there can be pumping noise even if there is zero adiabatically pumped charge. The pumping noise also indicates the respective direction of the bias-induced current and the pumping current.
\end{abstract}


\maketitle
\section{Introduction}

Transport through mesoscopic systems, such as quantum dots, can be achieved through a periodic time-dependent modulation of the system. This is referred to as \textit{pumping}. In the regime of adiabatic pumping, the modulation is slow, such that the system can \textit{almost} immediately follow the driving.  A directed transport through quantum dots in the absence of a stationary bias (or, alternatively, also an additional pumping contribution on top of or opposite to transport due to a stationary bias) is then possible only if there  are at least two independent pumping parameters at work.~\cite{Buttiker94,Brouwer98,Zhou99,Moskalets01,*Moskalets02a,Entin02,Avron04}

Adiabatic charge pumping offers the possiblity of a controlled emission of one charge per cycle, useful as a quantum standard for the current or as a coherent particle source for quantum operations.~\cite{Pothier92,Fletcher03,*Ebbecke05,Leek05,Blumenthal07,Feve07,Haack11,Pekola12} In addition to the current signal, knowledge about the noise created during a pumping cycle is vitally important, for instance in metrology to investigate the quantization limitations of the pump,~\cite{Andreev00,*Makhlin01,Avron01,Maire08} and also as a tool to detect signatures for emission of coherent electron packets.~\cite{Olkhovskaya08,Mahe10,*Bocquillon12} 

Recently there has also been strong interest in pumping due to quantum-interference effects,~\cite{Switkes99,Watson03,Mottonen08,Giazotto11,Gasparinetti12} where the pumped charge is not quantized. In this context it has been shown that the pumped charge may be used as a spectroscopic tool that reveals features beyond time-independent measurements.~\cite{Splettstoesser06,Reckermann10,Calvo12}

\begin{figure}
\includegraphics[width=0.3\textwidth]{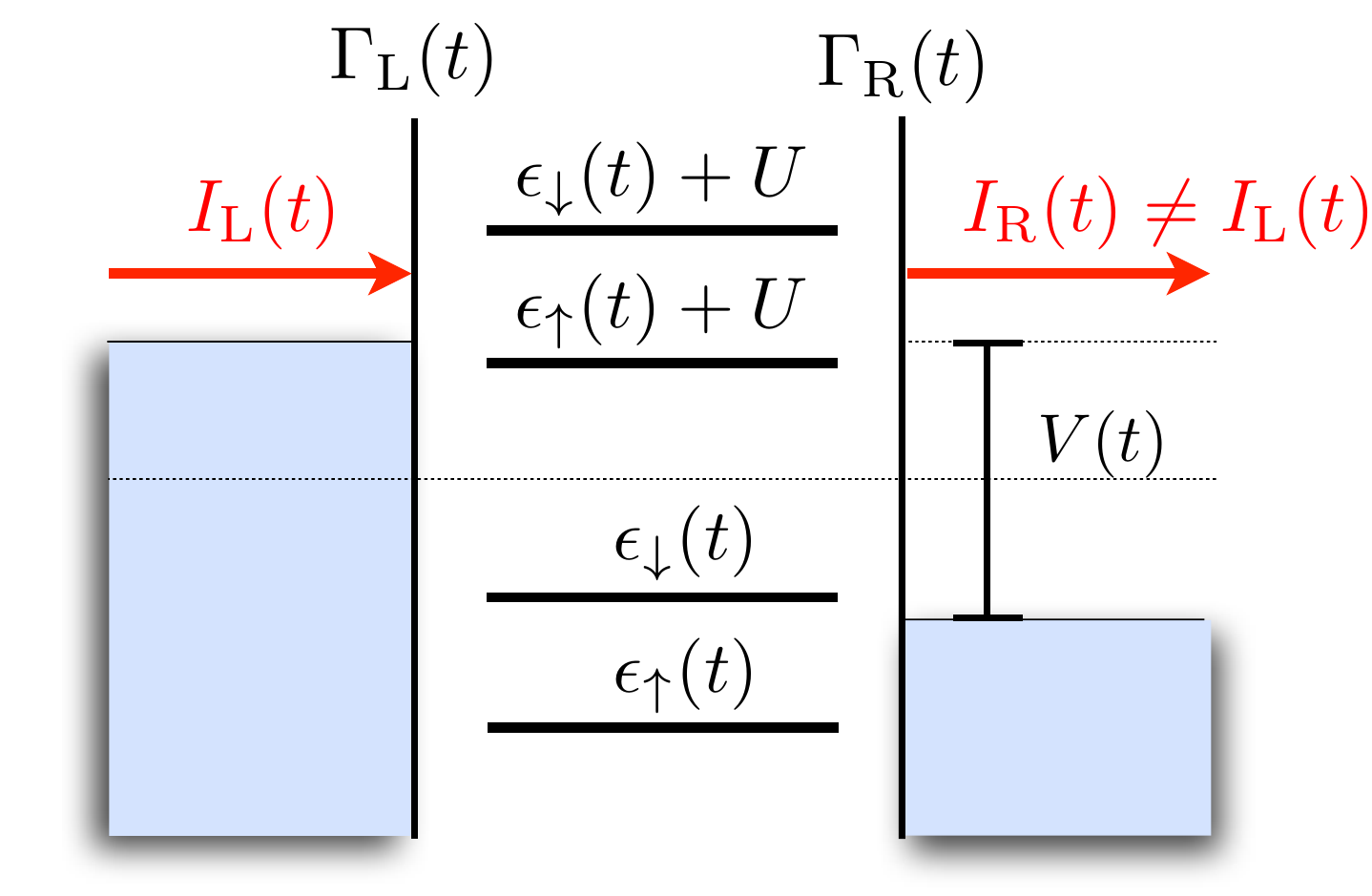}
\caption{Energy diagram of a single-level quantum dot attached to two leads via tunneling barriers. A symmetric voltage bias (with respect to the equilibrium chemical potential, see horizontal line) can be applied across the system and the energy level, the bias and the coupling strength may be chosen time dependent. Time-dependent charging and uncharging of the system goes along with a time-dependent difference between the currents flowing into the leads. 
}\label{fig_model}
\end{figure}

Another powerful tool for an extended transport spectroscopy is the transport noise. In this manuscript, we aim to identify additional spectroscopy features in the pumping noise. 
In the time-independent case with an external bias, there are two system-intrinsic noise contributions, namely thermal (Johnson-Nyquist) and shot noise, where the latter can help to determine the nature of the charge carriers and their statistics. Various theoretical and experimental works dealt with the study of noise in stationary transport.~\cite{Blanter00,Saminadayar97,*dePicciotto97, Oberholzer02,Braggio03,Beenakker03,Cottet04,Braggio06,Kiesslich07,Feldman07,Sothmann12,Basset12,Kumar12,Mueller12} For a time-dependently driven system an additional pumping noise emerges.~\cite{Moskalets08} For different types of time-dependent driving, noise and the full counting statistics has been investigated theoretically,\cite{Moskalets02b,Camalet03,Moskalets04b, Moskalets07,Kambly09,Safi11,Hammer11,*Vanevic12} and has  been proposed as a measure of quantum effects.~\cite{Lesovik93, Romito03, *Romito04,Flindt04,Olkhovskaya08,Devillard08,Albert10} The theoretical investigation of finite-frequency correlation functions in quantum dots driven by a time-dependent gate contacted to a single lead\cite{Mora10,Lee11,Nigg06} reveal information on the charge relaxation resistance.~\cite{Buttiker93,Kashuba12} Recently, also experiments measuring noise in time-dependently driven systems were performed.~\cite{Mahe10,*Bocquillon12,Gabelli08,*Grenier11,*Parmentier12,*Gabelli13}

When considering quantum pumps realized in small quantum dots, Coulomb interaction between electrons on the dot becomes important, due to the small capacitance of strongly confined systems. The impact of Coulomb interaction on the transport behavior of adiabatic pumps has recently been studied in various setups and regimes.\cite{Citro03,Aono04,Splettstoesser05,Brouwer05,Sela06,Arrachea08,Fioretto08,Hernandez08,Cavaliere09,Winkler09,*Hiltscher10} In particular, it has been shown that adiabatic quantum pumps can reveal characteristic fingerprints of the electron-electron interaction, such as interaction-induced renormalization effects,\cite{Splettstoesser06} and the impact of interactions on the relaxation rates of the dot charge.\cite{Reckermann10} Also the noise in stationary systems is sensitive to Coulomb interaction effects, see, e.g., Ref.~\onlinecite{Thielmann05} where super-Poissonian shot noise is observed in a Coulomb blockade regime due to inelastic spin-flip processes. The study of the noise in time-dependently driven quantum dots with strong electron-electron interaction has not been addressed so far. Here, we aim to close this gap, and include the Coulomb interaction with arbitrary strength.  Based on new noise features, we extend the possibilities of spectroscopy and find signals that give evidence of the underlying pumping mechanism.

In this paper we derive a formalism to compute the zero-frequency noise for adiabatically driven quantum systems, based on a real-time diagrammatic approach that enables us to include an arbitrary Coulomb interaction strength and non-equilibrium due to a bias and due to time-dependent driving.\cite{Konig96a, Konig96b} This general real-time formalism has  previously been the basis for the development of a method to calculate the zero-frequency noise in a strongly interacting quantum dot system driven by a \textit{stationary} bias\cite{Thielmann05} and for the development of a method to calculate charge and spin transport in quantum dot systems with a slow time-dependent driving.\cite{Splettstoesser06,Splettstoesser08a} 

We implement the formal results of the method developed here to study the pumping noise for the simplest generic model, a single-level quantum dot coupled to two leads, taking into account Coulomb interaction, for a wide range of possible time-dependent parameters. The quantum dot system is shown in Fig.~\ref{fig_model}. We are able to present analytic expressions for all driving regimes, allowing for a detailed analysis of the origin of the various noise contributions.

We start our discussion by investigating  the pumping noise when the non-equilibrium enters purely via the time-dependent driving. We recover a time-averaged fluctuation-dissipation relation~\cite{Moskalets04b,Moskalets08} and study deviations from it in the presence of both interaction and time-dependent driving. From the correction to the Fano factor due to pumping, we can (independent of the choice of pumping parameters) extract signatures of system-specific parameters, such as coupling asymmetry, as well as the level position with respect to the electron-hole symmetric point.

We, furthermore, consider the case of having an additional, possibly time-dependent, external bias. For a certain pumping prescription we find in the low-bias regime that the pumping noise is finite even if the time-averaged pumping current, namely its dc component, vanishes. We show how the \textit{time-averaged}  (i.e., zero-frequency) pumping noise feature provides information about the \textit{time-resolved} pumping current (i.e. the current at every instant of time). That is, a finite pumping noise is present if there is a finite time-resolved pumping current.

This manuscript is structured as follows. We start by introducing the model and outline the derivation of the noise formulae in Sec.~\ref{sec_model}. In this section we explicitly derive the technique to compute the zero-frequency noise. The results found for the pumping noise in a single level quantum dot, based on this technique, are discussed subsequently; the discussion is divided into two parts. The case of
pure pumping (no bias) is shown in Sec.~\ref{sec_results} and the results for pumping in presence of a finite bias are discussed in Sec.~\ref{sec_results_bias}. The conclusion in Sec.~\ref{sec_conclusion} is followed by several appendices detailing the derivation of the pumping noise.
There we introduce the treatment of slow time-dependences in general and for the zero-frequency noise in particular in Appendices~\ref{appendix_adiabatic_laplace} and~\ref{appendix_noise_adiabatic}. The real-time diagrammatic technique is explained in Appendix~\ref{appendix_diagrammatic_rules}.  Appendix~\ref{appendix_decaying_propagator} provides additional terms needed to evaluate the noise, while Appendix~\ref{appendix_analytic} contains some analytic expressions that are not shown in the main text.
\subsection{Current and noise in a quantum pump}\label{sec_definitions}

In this manuscript we address the current noise in an adiabatic quantum-dot pump, in which a current is induced by a slow \textit{time-dependent, periodic modulation} of the system parameters. The expectation value of the resulting pumping current is time-dependent  and fulfills the condition $\sum_\alpha I_\alpha(t)=-e\frac{d}{dt}\langle\hat{n}\rangle(t)$, with the currents, $I_\alpha(t)$, into the left and right leads, $\alpha=\mathrm{L,R}$,  and the charge in the (quantum dot) system, $e\langle\hat{n}\rangle(t)$ 
($e<0$ being the charge of the electron).  When at least two parameters of the 
system are varied in time, charge can be pumped through the system in every cycle. In this case, as well as  when an additional time-independent dc bias is applied, the \textit{time-averaged} current (namely the dc component of the current) is, in general, finite. The dc current through the system is defined as
\begin{equation}
\bar{I}=\int_0^\tau \frac{dt}{\tau}I(t)=\int_0^\tau \frac{dt}{\tau}\frac{1}{2}\left(I_\mathrm{L}(t)-I_\mathrm{R}(t)\right)\ ,
\end{equation}
where $\tau=2\pi/\Omega$ is the period and $\Omega$ the driving frequency. We are interested in the quasi stationary situation with a periodic modulation, where the expectation value of the charge on the dot is conserved after one period. Therefore, the time-averaged current $\bar{I}$ is equal to $\bar{I}_\mathrm{L}=-\bar{I}_\mathrm{R}$ due to charge conservation.

The current flowing through the pump is in general noisy, due to thermal fluctuations and due to non-equilibrium fluctuations. We are interested in the noise as a spectroscopy tool and as a measure of the precision of the charge pump. The time-resolved current noise can be defined in the symmetrised form as
\begin{equation}\label{eq_noise_two_times}
S_{\alpha\beta}\left(t,t'\right)=\left\langle \delta \hat{I}_{\alpha}\left(t\right)\delta \hat{I}_{\beta}\left(t'\right)+\delta \hat{I}_{\beta}\left(t'\right)\delta \hat{I}_{\alpha}\left(t\right)\right\rangle\ ,
\end{equation}
where the operator $\delta \hat{I}_{\alpha}\left(t\right)=\hat{I}_{\alpha}\left(t\right)-\langle \hat{I}_{\alpha}\left(t\right)\rangle$ measures the deviation of the current in lead $\alpha$ from its mean value. For a time-independent Hamiltonian, this object can be written as a function of the time difference, $t-t'$. For systems that are subject to an external time-dependent driving it depends in general on two times. We are interested in the noise from a long-time measurement, namely the zero-frequency contribution to the noise spectrum
\begin{eqnarray}\label{eq_zero_freq_noise}
S_{\alpha\beta} & = & \int_{0}^{\tau}\frac{dt}{\tau}\int_{-\infty}^{\infty}d(t-t')S_{\alpha\beta}\left(t,t'\right)\\\label{eq_zero_freq_noise_2}
& =: & \int_{0}^{\tau}\frac{dt}{\tau} S_{\alpha\beta}\left(t\right)\ .
\end{eqnarray}
Note that for the general noise-power definition, the integral in $t$ is taken over a large measuring time interval.~\cite{Blanter00} In the stationary limit, the integrand $S_{\alpha\beta}\left(t\right)=S_{\alpha\beta}\left(0\right)$ is time-independent. However, in the case of periodic driving, $S_{\alpha\beta}\left(t\right)$ as defined in Eq.~(\ref{eq_zero_freq_noise_2}) is periodic in $t$, and the integration interval can be reduced to one driving period $\tau$. We will in the following calculate the zero-frequency noise in time-dependently driven quantum dots.

\section{Model and Formalism}\label{sec_model}

\subsection{Model Hamiltonian}\label{sec_hamiltonian}

As a model system for the study of the pumping noise we consider a single-level quantum dot coupled to two leads $\alpha=\text{L},\text{R}$, as shown in Fig.~\ref{fig_model}. The quantum-dot system is  described by the Anderson model where certain parameters in the system are time dependent
\begin{align}
H&=\sum_{\alpha=\text{L},\text{R}}\sum_{k,\sigma=\uparrow,\downarrow}\left[\epsilon_{ k}+\mu_\alpha(t)\right]c_{\alpha k\sigma}^{\dagger}c^{}_{\alpha k\sigma}+\sum_{\sigma}\epsilon_\sigma\left(t\right)d_{\sigma}^{\dagger}d_{\sigma}^{} \nonumber\\ &+U\hat{n}_{\uparrow}\hat{n}_{\downarrow}+\sum_{\alpha=\text{L},\text{R}}\sum_{k\sigma}\left(\gamma_{\alpha}\left(t\right)d_{\sigma}^{\dagger}c_{\alpha k\sigma}+\text{h.c.}\right)\ .
\label{eq_hamiltonian}
\end{align}
The first term with the creation (annihilation) operator $c_{\alpha k\sigma}^{\dagger}$ ($c^{}_{\alpha k\sigma}$) describes electrons with spin $\sigma$ and momentum $k$  in the leads L and R, which can have different,  possibly time-dependent chemical potentials. The energy level on the dot is denoted as $\epsilon_{\uparrow,\downarrow}(t)=\epsilon(t)\mp\Delta/2$, and can be time dependent via driving by a gate. Eventually, we consider a finite magnetic field, leading to the Zeeman splitting $\Delta$. The number operator of electrons with spin $\sigma$ on the dot is $\hat{n}_\sigma=d_{\sigma}^{\dagger}d_{\sigma}^{}$, where $d_{\sigma}^{\dagger}$ ($d_{\sigma}^{}$) creates (annihilates) an electron with spin $\sigma$ on the dot. Coulomb interaction between electrons on the dot is included within the constant interaction model and is here parametrised by $U$. Electrons can tunnel from the leads onto the quantum dot, where the tunneling matrix element is given by the, possibly time-dependent, parameter $\gamma_{\alpha}(t)$. The tunnel coupling strength is characterized by the rates $\Gamma_\alpha(t,t')=2\pi\rho_\alpha\gamma_\alpha(t)\gamma^*_\alpha(t')$, with $\Gamma_\alpha(t)=\Gamma_\alpha(t,t)$ and the density of states of lead $\alpha$, $\rho_\alpha$. In the following we set $\hbar= 1$.

\subsection{Real-time diagrammatic approach for noise}\label{sec_diagram}

We want to derive the zero-frequency current noise for the above introduced model, taking into account the Coulomb interaction without any further approximation, and treating the case where the quantum system is weakly coupled to the reservoirs. This is accomplished in the following, where we write the zero-frequency current noise as well as the current itself based on a kinetic equation, which allows for an evaluation in terms of a non-equilibrium, real-time diagrammatic approach. 

In order to calculate these expectation values, information on the time-dependent density matrix of the system is needed. Since we are not interested in the specific time evolution of the states of the reservoirs, we trace out the degrees of freedom of the reservoirs and calculate the expectation values based on the knowledge of the reduced density matrix of the quantum dot and its time evolution. We are then left with a four-dimensional Hilbert space spanned by the dot states $|\chi\rangle\in\left\{|0\rangle,|\uparrow\rangle,|\downarrow\rangle,|\mathrm{d}\rangle\right\}$, namely the eigenstates of the Hamiltonian of the uncoupled quantum dot $\sum_{\sigma}\epsilon_\sigma\left(t\right)d_{\sigma}^{\dagger}d_{\sigma}^{}+U\hat{n}_{\uparrow}\hat{n}_{\downarrow}$. The state associated with an empty dot has energy $E_0=0$, the energy of the singly occupied dot with spin $\sigma=\uparrow,\downarrow$ is given by $E_\sigma=\epsilon_\sigma$, while the energy of the doubly occupied dot also includes the Coulomb interaction, $E_\mathrm{d}=2\epsilon+U$.

\begin{figure}[H]
\vspace{5mm}
\includegraphics[scale=1]{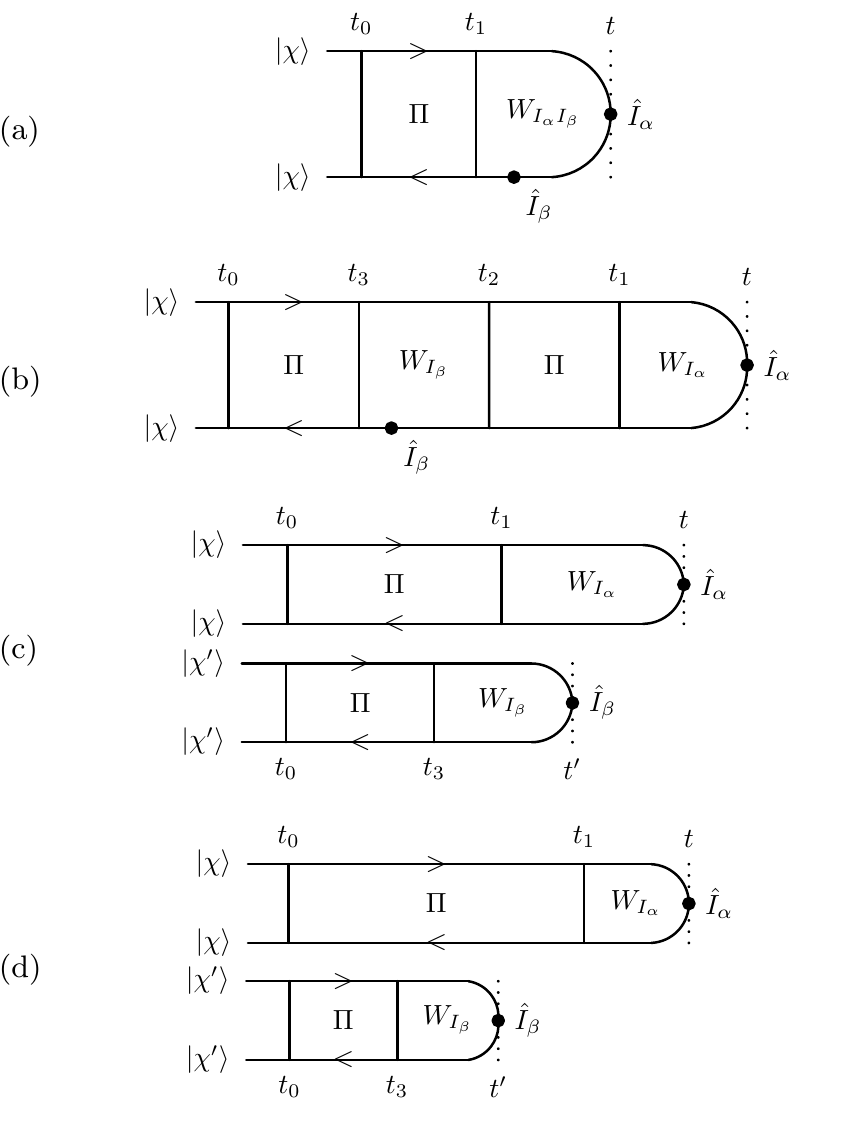}
\caption{Diagrams for the cases of having (a) both current operators in the same irreducible block, and (b) for having them on different blocks. Diagrams (c) and (d) depict the last contribution due to the product of current expectation values, where for (c) $t_1<t'$ and for (d) $t_1>t'$. 
\label{fig_contour}}
\end{figure}

We assume that the quantum dot and the reservoirs are decoupled at some initial time $t_0$; since we are not interested in transient effects, we will later set this initial time to $-\infty$, meaning that it is far away from the measuring time. 
The reduced density matrix of the dot at a later time $t$ is connected to the density matrix at time $t_0$ via  the propagator $\Pi(t,t_0)$.
In general, the reduced density matrix and the propagator are tensors of rank 2 and 4, respectively.
In the following, however, we concentrate on the time evolution of the diagonal elements of the reduced density matrix, summarized in the vector $P(t)=(P_0(t),P_\uparrow(t),P_\downarrow(t),P_\mathrm{d}(t))$. This is justified by the fact that in the model under consideration the dynamics of diagonal and off-diagonal elements of the reduced density matrix do not couple to each other due to charge and spin conservation in tunneling processes.
Therefore, we write
\begin{equation}
P(t) = \Pi(t,t_0)P(t_0)\ ,
\end{equation}
where the propagator $\Pi(t,t_0)$ is a four by four matrix connecting the vectors $P(t)$ and $P(t_0)$.
In order to capture the non-equilibrium time-evolution, we introduce the Keldysh contour. With this contour the time line, e.g., for the evolution of $P$ from $t_0$ to $t$ can be depicted by an upper (lower) time contour, representing the forward (backward) part of the time evolution, indicated by arrows. Such a contour is shown for the calculation of the noise, which will be introduced later, in Fig.~\ref{fig_contour}. The propagator fulfills a Dyson equation of the form
\begin{equation}\label{eq_Dyson_eq_time}
\Pi\left(t,t_0\right)=\mathbf{1}+\int_{t_0}^tdt_1\int_{t_0}^{t_1}dt_2W\left(t_1,t_2\right)\Pi\left(t_2,t_0\right)\ ,
\end{equation}
with the kernel $W$ as the self energy, which describes all processes coupling the quantum dot and the leads. The time evolution of the reduced density matrix can then be written in terms of a kinetic equation~\cite{Konig96a,Konig97}
\begin{equation}
\frac{d}{dt}P(t)=\int_{-\infty}^{t}dt'W(t,t')P(t')\ ,
\label{eq_kineq}
\end{equation}
where the kernel $W(t,t')$ accounts for transitions between different reduced density matrix elements due to tunneling.  
It
can be computed diagrammatically; 
we here use a diagrammatic language,~\cite{Konig98} where  the process of an electron hopping between dot and reservoir is represented as a tunneling vertex on the Keldysh contour. These vertices are connected via tunneling lines, which are contractions stemming from the tracing out of the lead degrees of freedom (due to Wick's theorem, there is always a pairwise connection of vertices). In this context, the kernel, i.e., the self energy of the Dyson equation~(\ref{eq_Dyson_eq_time}), can be regarded as the sum of all irreducible diagrams, namely those parts of the contour where by any vertical cut one crosses a tunneling line, see Appendix~\ref{appendix_diagrammatic_rules} for more details.

With this formalism one can treat the time-dependent expectation value of the current, $I_\alpha(t)=\langle \hat{I}_\alpha(t)\rangle$, by a similar equation as Eq.~(\ref{eq_kineq}). The current operator $\hat{I}_\alpha$ simply acts as an external vertex which can be connected to the previously introduced tunneling vertices, via tunneling lines. We therefore introduce the object $W_{I_\alpha}$ which represents the sum of all irreducible diagrams including the current operator as an external vertex. Based on this, the current expectation value can be written as
\begin{equation}\label{eq_def_current}
I_{\alpha}\left(t\right)=\frac{e}{2}\text{e}^{\text{T}}\int_{-\infty}^{t}dt'W_{I_{\alpha}}\left(t,t'\right)P\left(t'\right)\ ,
\end{equation}
with $\text{e}^{\text{T}}=(1,1,1,1)$, being a vector representation of the trace operator acting on the reduced Hilbert space.
The current kernel $W_{I_{\alpha}}$ as appearing in Eq.~(\ref{eq_def_current}) can be directly derived from the kernel $W$ by use of the accordingly modified diagrammatic rules shown in Appendix~\ref{appendix_diagrammatic_rules}.

We are in the following interested in the zero-frequency noise. In order to express the zero-frequency noise based on the time evolution of the current operators on the Keldysh contour, we concentrate on the object $S_{\alpha\beta}(t)$. This is the integrand of the expression for the zero-frequency noise, given in Eq.~(\ref{eq_zero_freq_noise}). Here we deal with an expectation value that contains two current operators at two different times $t$ and $t'$. As in the previous case of the current expectation value, we can treat the current operators as external vertices connected via tunneling lines. However, the placing of two external current vertices on the Keldysh contour allows for more possibilities of connections which need to be taken care of separately, as we show in the following (see also Fig.~\ref{fig_contour}).

We distinguish the two cases where the time argument of the current operator $\hat{I}_\alpha$ is larger than the time argument of $\hat{I}_\beta$, namely $t>t'$, yielding the three integral expressions of the following formula, and the opposite situation $t'>t$, where the current $I_\beta$ is taken at a later time and which we obtain by exchanging the indices $\alpha$ and $\beta$, 
\begin{widetext}
\begin{align}
S_{\alpha\beta}\left(t\right)=&\frac{e^{2}}{2}\text{e}^\text{T}\lim_{t_0\rightarrow-\infty}\biggl[\int_{t_{0}}^{t}dt_{1}W_{I_{\alpha}I_{\beta}}\left(t,t_{1}\right)P\left(t_{1}\right)+\int_{t_{0}}^{t}dt_{1}\int_{t_{0}}^{t_{1}}dt_{2}\int_{t_{0}}^{t_{2}}dt_{3}W_{I_{\alpha}}\left(t,t_{1}\right)\Pi\left(t_{1},t_{2}\right)W_{I_{\beta}}\left(t_{2},t_{3}\right)P\left(t_{3}\right)\nonumber\\&-\int_{t_{0}}^{t}dt_{1}\int_{t_{0}}^{t}dt'\int_{t_{0}}^{t'}dt_{3}W_{I_{\alpha}}\left(t,t_{1}\right)P\left(t_{1}\right)\otimes \text{e}^\text{T}W_{I_{\beta}}\left(t',t_{3}\right)P\left(t_{3}\right)\biggr] +(\alpha\leftrightarrow\beta)\ .\label{eq_noisepart_det}
\end{align}
\end{widetext}
We discuss in the following the three different integral expressions occurring for $t>t'$, examples of which are depicted in Fig.~\ref{fig_contour}. 

The first term of Eq.~(\ref{eq_noisepart_det}) deals with the possibility to place both current operators in the same irreducible part on the Keldysh contour, captured by the object $W_{I_{\alpha}I_{\beta}}$, see Fig.~\ref{fig_contour}~(a). The operator $\hat{I}_\alpha$ occurs at time $t$, which is the latest time within $S_{\alpha\beta}\left(t\right)$, see Eq.~(\ref{eq_noisepart_det}), and is placed at the turning point of the Keldysh contour.
The second current operator $\hat{I}_\beta$ has no fixed position within the object $W_{I_{\alpha}I_{\beta}}$. We here show the example where the current operator $\hat{I}_\beta$ is placed on the lower contour, however also the diagrams where $\hat{I}_\beta$ is placed on the upper contour contribute to the situation where $t>t'$.

In the second term the current operators are in \textit{different} irreducible parts, which are connected via the propagator $\Pi$ of the system, see Fig.~\ref{fig_contour}~(b). Also here, we fix the operator $\hat{I}_\alpha$ at the contour turning point at time $t$, whereas $\hat{I}_\beta$ can be placed anywhere on the contour (in Fig.~\ref{fig_contour}~(b), the lower contour part was chosen).

The last term accounts for the contribution due to the product of two separate current expectation values at different times, where the tensor product occurs as the product of two terms which can be depicted on separate Keldysh contours, see Fig.~\ref{fig_contour}~(c) and (d). The times, $t$ and $t'$, at which the currents, $\hat{I}_\alpha$ and $\hat{I}_\beta$, are taken can occur in different realizations: The time $t'$ can lie within the time interval set by the Kernel $W_{I_\alpha}(t,t_1)$, this is shown in the example Fig.~\ref{fig_contour}~(c). Note that, furthermore, the relation between the times $t_1$ and $t_3$, the times at which the current kernels start, is not fixed; in Fig.~\ref{fig_contour}~(c) we show as an example the case where $t_1>t_3$.
Alternatively, the time $t'$ can lie outside the time interval set by the kernel $W_{I_\alpha}(t,t_1)$, i.e., $t'<t_1$, see Fig.~\ref{fig_contour}~(d). In this case the third integral expression has a similar structure as the second one, depicted in Fig.~\ref{fig_contour}~(b). We use this fact for the evaluation of the zero-frequency noise, extensively discussed in the Appendix~\ref{appendix_noise_adiabatic}, where we combine these two contributions to one. 


Now we take the initial time $t_0$ (before which we suppose the system and the leads to be decoupled) to the limit $t_0\rightarrow-\infty$. The reason why we did not take this limit previously, is that the different integral expressions in Eq.~(\ref{eq_noisepart_det}) do not converge separately for this limit. This is due to the fact that the propagator $\Pi\left(t_{1},t_{2}\right)$ does not have a finite support, meaning that the propagator does not vanish in the limit $t_{1}-t_{2}$ going to infinity.


The discussed object, $S_{\alpha\beta}\left(t\right)$, is periodic in time and it, furthermore, enables us  to perform in a subsequent step a systematic
adiabatic expansion of the noise with respect to time $t$. The zero-frequency noise has before  been studied based on a real-time diagrammatic approach for a \textit{time-independent} system in Ref.~\onlinecite{Thielmann05}. Here, we are interested in considering the explicit time-dependence  arising from the pumping cycle affecting the zero-frequency noise. The general expression given in Eq.~(\ref{eq_noisepart_det}) is the starting point of our considerations. The treatment of the time dependence is presented in the following Secs.~\ref{sec_adiabatic_current} and~\ref{sec_adiabatic}.

\subsection{Adiabatic expansion of the current}\label{sec_adiabatic_current}

We are interested in a situation where a time-dependent modulation of the quantum dot system induces a current flowing through it, such as in a quantum pump. This time-dependent modulation can for example be realised by time-dependent gate voltages applied to the quantum dot. The different system parameters, which we choose to be possibly time-dependent are  indicated in the Hamiltonian in Eq.~(\ref{eq_hamiltonian}). In the case of \textit{adiabatic pumping}, which we focus on, this time dependence is taken to be periodic and slow enough such that the system can always almost follow the time-dependent modulation. This means that the time-scale set by the driving is much larger than the lifetime of electrons in the system. For a small-amplitude  modulation of the single-level quantum dot investigated here, this reduces to the condition $\Omega\ll\Gamma$; we will discuss more general conditions later in this section.


We aim to write the noise, Eq.~(\ref{eq_noisepart_det}), for the limit of slow driving. To this end we will perform an expansion in the driving frequency, taking into account the zeroth and the first order in $\Omega$. The zeroth order, referred to as the instantaneous contribution, describes the situation where the system can be taken to always follow the parameter modulation. This means that all parameters are frozen to time $t$.
An "adiabatic correction", in first order in $\Omega$, accounts for the actual delay with respect to the instantaneous solution. This contribution in first order in the frequency is responsible for the pumping current and the pumping noise. In the following we indicate the  contribution in zeroth order of the frequency expansion with the superscript $(i)$ and we denote the first-order correction in frequency by $(a)$.


As a basis to address the adiabatic expansion of the zero-frequency noise in orders of the driving frequency, we first perform the expansion of the kinetic equations and the current, following the lines of Ref.~\onlinecite{Splettstoesser06}. In this spirit, the kinetic equation for the adiabatic expansion of the reduced density matrix of the dot, $P(t)\rightarrow P^{(i)}_t+P^{(a)}_t$, starting from Eq. (\ref{eq_kineq}), is given by
\begin{align}
0&=\left\{WP\right\}^{(i)}_t\label{eq_master_instantaneous}\\\label{eq_master_adiabatic}
\frac{d}{dt}P^{(i)}_t&=\left\{WP\right\}^{(a)}_t\ .
\end{align}
The instantaneous contribution and the adiabatic correction to the reduced density matrix are found as solutions of these two sets of equations, which contain the instantaneous contribution and the adiabatic correction  to the kernel, $W\rightarrow W^{(i)}+W^{(a)}$, as well. We use a short notation with brackets, which is generally defined for two arbitrary objects $A$ and $B$ (be it any of the kernels, the density matrix, or also products of several objects) as
\begin{align}
\left\{AB\right\}_t^{(i)}&=A_t^{(i)}B_t^{(i)}\\\label{eq_adiabatic_bracket}
\left\{AB\right\} _{t}^{\left(a\right)}&=A_{t}^{\left(i\right)}B_{t}^{\left(a\right)}+A_{t}^{\left(a\right)}B_{t}^{\left(i\right)}+\partial A_{t}^{\left(i\right)}\dot{B}_{t}^{\left(i\right)}\ .
\end{align}
This expression can be readily derived from a general expression for the frequency expansion of a convolution, as previously done by Kashuba \textit{et al.} Ref.~\onlinecite{Kashuba12}, see also Appendix~\ref{appendix_adiabatic_laplace}.  In these equations, the Laplace transforms of the instantaneous  kernel and its adiabatic correction, $W_t^{(i/a)}(z)={\int_{0}^{\infty}d(t-t')e^{-z(t-t')}W_t^{(i/a)}(t-t')}={\int_{-\infty}^{t}dt'e^{-z(t-t')}W_t^{(i/a)}(t-t')}$ occur, where the short notation without the $z$-argument denotes the limit $W_t^{(i/a)}:=\lim_{z\rightarrow0^+}W_t^{(i/a)}(z)$
and $\partial W_t^{(i)} := \lim_{z\rightarrow0^+} [\partial W_t^{(i)}(z)/\partial z]$. 

The kernel $W_t^{(i)}$ can be constructed in a straightforward manner by taking the kernel of the stationary system and inserting the time dependence parametrically. This is indicated by the subscript $t$. The adiabatic correction to the kernel contains the effect of the time-dependent parameters changing in time along the Keldysh contour in first order in $\Omega$ and is obtained according to the rules given in Appendix~\ref{appendix_diagrammatic_rules}, see also Ref.~\onlinecite{Splettstoesser06}. The fact that the Kernel has a finite support in time enters this equation through 
the derivative $\partial W_t^{(i)}$, which is a non-markovian feature.
\footnote{For the general case where $A$ is the convolution of several objects, $\partial A$ is to be understood as the frequency derivative of the convolution at $z=0$. }

We can deduct from the above equations, that in order to justify the adiabatic expansion of the kinetic equation for a general setting, one has to take into account two relevant time scales. For one there is the dynamics of the density matrix, dominated by tunnelling processes due to coupling to the reservoirs. Since the effect of the tunnel coupling is incorporated in the kernel $W$, the relevant time scale is directly related to the inverse of the magnitude of $W$. Moreover, $W(t-t')$ has a finite support in time, which needs to be sufficiently small to validate the expansion in $\Omega$. The second time scale is hence related to the width of the kernel, which in turn relates to $\partial W$ and corresponds to the dynamics of the reservoir. In this spirit, the adiabatic expansion is justified if both the tunnel and the reservoir dynamics are much faster than the external driving.

Due to the slow parameter modulation, also the current through the system has an instantaneous contribution and a first-order-in-frequency, adiabatic correction. The first-order correction is exactly the pumping current, in which we are mainly interested. In analogy to the kinetic equation one can compute the current of the system by the respective adiabatic expansion of the current kernel and the reduced density matrix
\begin{align}\label{eq_Ii}
I_\alpha^{(i)}(t)&=\frac{e}{2}\text{e}^\text{T}\left\{W_{I_\alpha}P\right\}^{(i)}_t\\\label{eq_Ia}
I_\alpha^{(a)}(t)&=\frac{e}{2}\text{e}^\text{T}\left\{W_{I_\alpha}P\right\}^{(a)}_t\ .
\end{align}
The reduced density matrix, obtained from Eqs.~(\ref{eq_master_instantaneous}) and (\ref{eq_master_adiabatic}), enters the formula for the current expectation value. The current kernel contributions $W^{(i/a)}_{I_\alpha,t}$ and $\partial W^{(i)}_{I_\alpha,t}$ are constructed according to the rules in Appendix~\ref{appendix_diagrammatic_rules}. Importantly, if no bias voltage is applied, the instantaneous current is always zero. In this case, the pumping current, $I^{(a)}_\alpha$, becomes the dominant contribution. We are in the following interested in the situation of zero bias, namely the regime of \textit{pure pumping} as well as in the situation where a pumping current flows on top of an instantaneous current due to a non-equilibrium bias.

\subsection{Adiabatic expansion of the noise}\label{sec_adiabatic}

The main purpose of this paper is the calculation of the zero-frequency current noise of an adiabatic quantum-dot pump. The aim of this section is to present the zeroth-order and first-order contribution in the driving frequency to the zero-frequency noise, ${S}_{\alpha\beta}^{\left(i\right)}$ and ${S}_{\alpha\beta}^{\left(a\right)}$, going along with the pumping current presented in the previous section. For this purpose we expand the elements of the periodic function $S_{\alpha\beta}\left(t\right)$, given in Eq.~(\ref{eq_noisepart_det}), around the reference time $t$. 
In a first step we consider the instantaneous contribution to the different terms, only. The detailed derivation is given in Appendix~\ref{appendix_noise_adiabatic}. 
We obtain for the zeroth-order contribution in the driving frequency to this function 
\begin{equation}
\begin{split}
S_{\alpha\beta}^{\left(i\right)}\left(t\right)=\left(\frac{e^{2}}{2}\text{e}^\text{T}\left[\left\{W_{I_{\alpha}I_{\beta}}P\right\}_{t}^{\left(i\right)}+\left\{W_{I_{\alpha}}\overline{\Pi}W_{I_{\beta}}P\right\}_{t}^{\left(i\right)}\right]\right.\\-2\left.\left\{ \tilde{I}_{\alpha} I_{\beta}\right\}_t^{\left(i\right)}\right)+(\alpha\leftrightarrow\beta)\ ,
\end{split}\label{eq_noise_inst}
\end{equation}
where we describe the different elements in the following. The first term contains the kernel $W_{I_{\alpha}I_{\beta}}$ involving two current operators; the functional form of this expression is analog to the one found for the instantaneous current. The second term, containing current kernels from different parts of the contour is a combination of the second and part of the third expression of Eq.~(\ref{eq_noisepart_det}). Note that an object related to the propagator connects the two kernels~\cite{Thielmann04a,Thielmann05}
\begin{equation}\label{eq_dec_prop}
\overline{\Pi}\left(t,t'\right)=\Pi\left(t,t'\right)-P(t)\otimes\text{e}^\text{T}\ .
\end{equation}
This object has a finite support and is, hence, zero for large time differences, $t-t'\rightarrow\infty$. Its further properties and its evaluation starting from a Dyson equation are discussed in Appendix~\ref{appendix_decaying_propagator}. The important trait of $\overline{\Pi}$ in order to justify the adiabatic expansion, is that the width of $\overline{\Pi}$ needs to be sufficiently small. The relevant time scale determining the support of $\overline{\Pi}$ is equal to the time scale determining the dynamics of the density matrix. Hence, it scales with the inverse of the tunneling rate. Therefore, there is no new relevant time scale appearing with respect to the ones discussed before, and the same conditions that justify an adiabatic treatment of the kinetic equation, Eqs.~(\ref{eq_Ii}) and (\ref{eq_Ia}), justify likewise the expansion of the noise for small pumping frequencies $\Omega$. The condition to evaluate the instantaneous contribution to the object $\overline{\Pi}$ is given in Eq.~(\ref{eq_Pbar_inst}).


Finally the third term of Eq.~(\ref{eq_noise_inst}) contains the remaining part of the product of two current expectation values at different times. In this third integral expression, the object 
\begin{equation}\label{eq_tilde_current}
\tilde{I}_\alpha(t,t_2)=\frac{e}{2} \mathrm{e}^\mathrm{T}\int_{-\infty}^{t_2}dt_1W_{I_{\alpha}}(t,t_1)P(t_1)\ ,
\end{equation}
occurs, which is an object depending on two time-arguments and differs from the definition of the current, Eq.~(\ref{eq_def_current}), by the integration interval, such that $\tilde{I}_\alpha(t,t)=I_\alpha(t)$. The Laplace transform of this object is discussed in detail in Appendix~\ref{appendix_noise_adiabatic}. Its zeroth-order contribution in the driving frequency  for $z\rightarrow 0^+$ is given by $\lim_{z\rightarrow0^+}\tilde{I}^{(i)}_\alpha(t,z)=-\partial I_{\alpha}^{\left(i\right)}\left(t\right)$, where we introduced the abbreviation $\partial I_{\alpha}^{\left(i\right)}\left(t\right)=\frac{e}{2}\text{e}^\text{T}\partial W_{I_{\alpha},t}^{\left(i\right)}P_{t}^{\left(i\right)}$. 
 
Integrating the function $S_{\alpha\beta}^{(i)}$ of Eq.~(\ref{eq_noise_inst}) over one period in order to obtain the zero-frequency noise as defined in Eq.~(\ref{eq_zero_freq_noise}) delivers a time-averaged version of the stationary zero-frequency noise considered by Ref.~\onlinecite{Thielmann04a,Thielmann05}. We will show for instance in Sec.~\ref{sec_quantized_pumping}, that if the time-dependent driving of the system occurs with \textit{large amplitudes}, already the noise in zeroth order in the pumping frequency can deviate strongly from the one of the corresponding stationary system.

The important object for this paper is the pumping noise, namely the first-order in the pumping frequency correction to the noise,  arising from the slow driving. Only if there is a working pump, i.e., only if there are two time-dependent parameters enclosing a finite area in parameter space in one period  the pumping noise is non-zero. With the help of the notation using brackets, introduced in the context of the kinetic equation in the previous section, the elements of the pumping noise can be written as
\begin{equation}
\begin{split}
S_{\alpha\beta}^{\left(a\right)}\left(t\right)=\left(\frac{e^{2}}{2}\text{e}^\text{T}\left[\left\{ W_{I_{\alpha}I_{\beta}}P\right\} _{t}^{\left(a\right)}+\left\{ W_{I_{\alpha}}\overline{\Pi}W_{I_{\beta}}P\right\} _{t}^{\left(a\right)}\right]\right.\\-2\left.\left\{\tilde{I}_{\alpha}I_{\beta}\right\}^{\left(a\right)}_t\right)+\left(\alpha\leftrightarrow\beta\right)\ .\label{eq_noise_ad}
\end{split}
\end{equation}
That is, each bracket contributing to the noise is expanded in first order in $\Omega$, according to Eq.~(\ref{eq_adiabatic_bracket}).
When explicitly expanding these brackets, the adiabatic correction and the first derivative with respect to the Laplace variable of the object $\overline{\Pi}$ related to the propagator appear, in addition to the previously found adiabatic correction to the reduced density matrix and the current. The recipes to evaluate $\partial\overline{\Pi}^{(i)}_t$ and $\overline{\Pi}^{(a)}_t$ are given in Eqs.~(\ref{eq_dPbar_inst}) and~(\ref{eq_Pbar_ad}).

The adiabatic correction to the third term of Eq.~(\ref{eq_noise_inst}) contains again the object $\tilde{I}$. For the evaluation of the adiabatic correction of the bracket $\left\{\tilde{I}_{\alpha}I_{\beta}\right\}$ we refer to the Appendix~\ref{appendix_noise_adiabatic}, in particular to Eqs.~(\ref{eq_dItilde}) and~(\ref{eq_Itilde_a}).
 
\subsection{Perturbation expansion in the tunnel coupling}\label{sec_tunnel_exp}

In the following we are interested in quantum dots which are weakly coupled to the reservoirs. We therefore perform a perturbation expansion in the tunnel coupling $\Gamma$ on top of the adiabatic expansion performed above. This perturbation approach is a useful approximation as long as the condition $\Gamma\ll k_\mathrm{B} T$ is fulfilled, where $T$ is the temperature and $k_\mathrm{B}$ the Boltzmann constant. Within this perturbation expansion, the Coulomb interaction and the non equilibrium due to an applied bias can be treated without any further approximations. The kernels describing tunneling events between quantum dot and reservoirs are always expressions in at least first order in the tunneling coupling strength $\Gamma$. We write their expansion in general as
\begin{equation}
W_{x,t}^{(i/a)}\approx W_{x,t}^{(i/a,1)}+W_{x,t}^{(i/a,2)}+\ldots\ ,
\end{equation}
where the index $x$ stands for the different kinds of kernels, such as $W$, the current kernel $W_{I_\alpha}$, or $W_{I_\alpha I_\beta}$. While expressions containing the kernel only in the lowest-order expansion in the tunnel coupling are commonly referred to as the sequential tunneling contribution, we here go beyond this limit by taking into account quantum fluctuation effects up to second order in the tunnel coupling $\Gamma$. 

We perform a rigorous order-by-order $\Gamma$ expansion of the kinetic equations given in Eqs.~(\ref{eq_master_instantaneous}) \& (\ref{eq_master_adiabatic}),  the currents from Eqs.~(\ref{eq_Ii}) \& (\ref{eq_Ia}),  the zero-frequency noise from Eqs.~(\ref{eq_noise_inst})  \& (\ref{eq_noise_ad}).
Thus we find the instantaneous and adiabatic contributions to the reduced density matrix, the current and the zero-frequency noise in lowest and in next-to-lowest order in the tunnel-coupling strength $\Gamma$. The results are discussed in the remainder of this paper. It is known from previous calculations,~\cite{Konig98,Splettstoesser06} that the lowest-order contribution to the reduced density matrix in the instantaneous limit is the zeroth order in $\Gamma$, denoted by $P^{(i,0)}$, while the adiabatic correction starts in minus first order, $P^{(a,-1)}$. This is consistent with the fact that $\Omega/\Gamma\ll1$ in the adiabatic limit. The dominant contribution to the instantaneous current, $I^{(i,1)}(t)$, is a first-order in $\Gamma$ contribution, while the pumping current starts in zeroth order in $\Gamma$ but is proportional to the frequency, $I^{(a,0)}(t)$. Likewise, the perturbation expansion of the instantaneous zero-frequency current noise begins in first order in $\Gamma$, $S_{\alpha\beta}^{(i,1)}$, where the pumping noise contribution starts one order lower, $S_{\alpha\beta}^{(a,0)}$. In the remainder of this paper we apply these general formulas to the single-level quantum dot, introduced in Sec.~\ref{sec_model}, for a slow, time-dependent driving, in the zero- as well as in the finite-bias regime.

\section{Pumping noise}\label{sec_results}
In this section we want to study the zero-frequency noise of a single-orbital quantum dot, as introduced in Sec.~\ref{sec_hamiltonian}. We consider the case of zero bias voltage, where a finite pumped charge is due to the modulation of at least two time-dependent parameters. In the absence of a bias the instantaneous current is zero for all times. However, a finite charge can be pumped through the dot, i.e., the dc pumping current $\overline{I}^{(a)}$ is non-zero, if there is a finite phase difference between the two parameters. These parameters  can be the level position of the dot and the tunnel-coupling strengths to the different contacts, which can all be driven by externally applied gates. In the following we write these parameters as $X(t)=\bar{X}+\delta X(t)$, where $\bar{X}$ denotes the time average of $X(t)$.

\subsection{Instantaneous contribution to the noise}\label{sec_pumping_inst}

While the instantaneous \textit{current} always vanishes at zero bias, \textit{thermal noise} is present, even without the modulation of external gates. We start the study of the noise in the unbiased case by this instantaneous contribution and show to which extent this thermal noise reveals features of the time-dependent driving. In lowest order in the tunnel coupling, we find
\begin{equation}\label{eq_inst_noise_Gamma1}
S_{\mathrm{LR}}^{\left(i,1\right)}=-4e^{2}\int_{0}^{\tau}\frac{dt}{\tau}\frac{\Gamma_\mathrm{L}\Gamma_\mathrm{R}}{\Gamma}\frac{\lambda_c^{(1)}}{\Gamma}\left\langle \Delta \charge\right\rangle _{t}^{\left(i,0\right)}\ .
\end{equation}
Note that we can express the noise exclusively in quantities, which are local in time and which characterize the properties of the quantum dot.~\cite{Davies92} Namely, Eq.~(\ref{eq_inst_noise_Gamma1}) contains the relaxation rate of the charge on the dot in lowest order in the tunneling, $\lambda_c^{(1)}=\Gamma\left[1+f(\epsilon)-f(\epsilon+U)\right]$,\cite{Splettstoesser10,Contreras12} with the Fermi function $f(E)=1/[1+\exp(\beta E)]$ and $\beta=1/k_\mathrm{B}T$,  as well as 
the instantaneous charge variance on the quantum dot, defined as
\begin{equation}
\left\langle\Delta \charge\right\rangle_{t}^{\left(i,0\right)}=\left\langle \charge^{2}\right\rangle _{t}^{\left(i,0\right)}-\left[\left\langle \charge\right\rangle _{t}^{\left(i,0\right)}\right]^{2}\ ,
\end{equation}
with the definition $\langle \charge \rangle:=\langle n\rangle-1$, where $\langle n\rangle$ is the average charge on the dot. The quantity $\langle\charge\rangle$ represents an average electron-hole occupation, namely it is $1$ if the dot is doubly occupied, and $-1$ for an empty dot, i.e., there are two holes. Otherwise, i.e., for one electron on the dot, $\charge$ is zero. Note that $\langle\Delta\charge\rangle=\langle\Delta n\rangle$. We choose this representation since the quantity $\langle\charge\rangle$ appears in noise expressions much more often than $\langle n\rangle$ alone. The reason for this is that the noise is related to charge fluctuations. We will see in the following, that the noise reveals interesting symmetries with respect to the electron-hole symmetric point. The explicit result for the quantum-dot charge variance can be given as
\begin{equation}
\langle\Delta \charge\rangle_t^{(i,0)}=\left(1-\frac{\lambda_c^{(1)}}{2\Gamma}\right)\left(1-\left[\langle \charge\rangle_t^{(i,0)}\right]^2\right)\ ,
\end{equation}
with $\langle \charge\rangle_t^{(i,0)}=2f(\epsilon)/\left[1+f(\epsilon)-f(\epsilon+U)\right]-1$.

\begin{figure}
\includegraphics[scale=1]{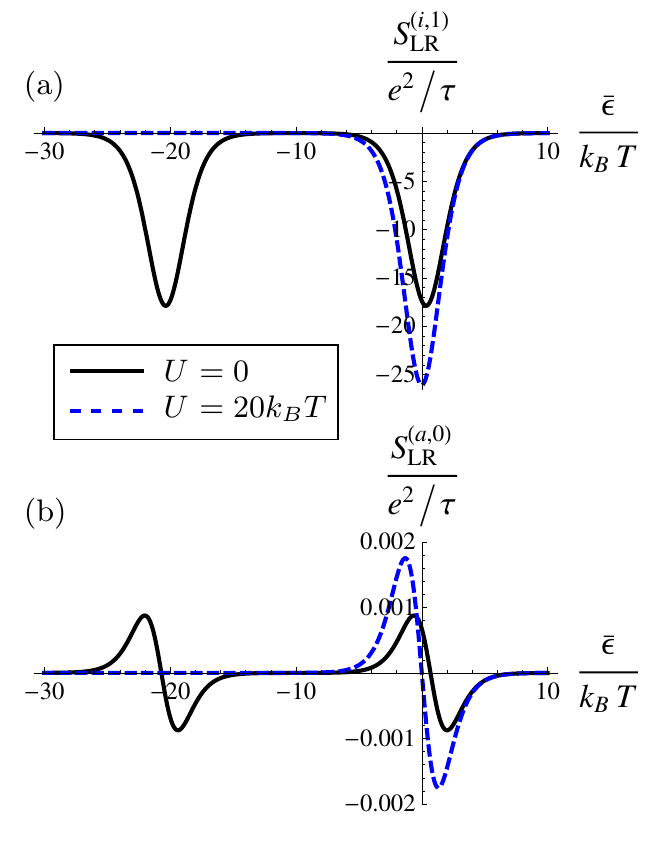}
\caption{Plot for (a) the instantaneous and (b) the adiabatic zero-frequency noise as a function of the mean energy level $\overline{\epsilon}$. The graphs are drawn for $U=0,20k_\mathrm{B}T$ (blue dashed, black solid). The left and right tunnel coupling rates are $\Gamma_\mathrm{L}=0.3\Gamma$ and $\Gamma_\mathrm{R}=0.7\Gamma$. The other parameters are $\Gamma=0.5k_\mathrm{B}T$, the pumping amplitudes $\delta\epsilon,\delta\Gamma_\text{L}=0.1\Gamma$, the phase difference between the pumping parameters $\varphi=\pi/2$, and $\Omega=0.1\Gamma$.}
\label{fig_si_sa_nobias}
\end{figure}

While the above discussion was valid for arbitrary pumping parameters, we now focus on the case of time-dependent energy level and left tunneling rate, $\epsilon=\overline{\epsilon}+\delta\epsilon\sin(\Omega t)$ and $\Gamma_\text{L}=\overline{\Gamma}_\text{L}+\delta\Gamma_\text{L}\sin(\Omega t+\varphi)$, where pumping amplitudes are small, $\delta\Gamma_\mathrm{L},\delta\epsilon\ll\bar{\Gamma}$. The result for the instantaneous zero-frequency noise is shown in  Fig.~\ref{fig_si_sa_nobias}~(a). The instantaneous cross-correlation is always negative, as required.\cite{Blanter00} For small amplitudes it shows a resonant contribution at $\bar{\epsilon}\approx0$, when the addition energy of adding an electron to the empty dot is equal to the electrochemical potential, and at $\bar{\epsilon}\approx-U$, when a second electron can be added. These two contributions coincide, when the Coulomb interaction is zero (see blue, dashed line). One can observe, that the single peak for $U=0$ is not the sum of the two peaks for finite $U$, but is reduced by a factor of roughly $0.7$. In Eq.~(\ref{eq_inst_noise_Gamma1}) we see that this arises from the fact that the tunneling rate prefactor is modified by the Coulomb interaction through $\lambda_c^{(1)}/\Gamma$. For $U\neq0$, there is an increase of the rates, since $\lambda_c^{(1)}/\Gamma>1$ when $\overline{\epsilon}$ is within the range from $-U$ to $0$. This leads to an effective increase of the total current fluctuations for finite $U$.

Irrespective of the choice of pumping parameters we can state that the instantaneous part of the noise is a pure thermal noise, since no bias is applied. Equation~(\ref{eq_inst_noise_Gamma1}) can be understood as a \textit{time-averaged} fluctuation dissipation theorem for it can be related to the instantaneous linear conductance as 
\begin{equation}\label{eq_FDT}
S_{\mathrm{LR}}^{\left(i\right)}=-4k_{B}T\int_{0}^{\tau}\frac{dt}{\tau}G^{\left(i\right)}_t\ ,
\end{equation}
where the linear conductance is defined as
\begin{equation}
G=\left.\frac{\partial I}{\partial V}\right|_{V=0}\ .
\end{equation}
This result holds also in next order in the tunnel coupling, with  corrections to the linear conductance of the form\cite{Konig99, Splettstoesser12}
\begin{equation}
G^{(2)}=\sigma \frac{\partial G^{(1)}}{\partial{\epsilon}}+\sigma_{\Gamma} G^{(1)} +G^{(\text{cot})}\ ,
\label{eq_G_static}
\end{equation}
with the lowest-order conductance $G^{(1)}$, which can be directly extracted from Eq.~(\ref{eq_inst_noise_Gamma1}),  and 
 the definitions
\begin{align}
\sigma&=\phi(\epsilon+U)-\phi(\epsilon)\\
\sigma_\Gamma&=\left(\frac{2}{U}\sigma-\phi'(\epsilon)-\phi'(\epsilon+U)\right)\frac{1-f(\epsilon)-f(\epsilon+U)}{1-f(\epsilon)+f(\epsilon+U)}\ .
\end{align}
We defined $\phi(\omega)=\frac{\Gamma}{2\pi}\text{Re}\Psi\left(\frac{1}{2}+i\frac{\beta\omega}{2\pi}\right)$ and $\phi'(\omega)=\frac{\partial}{\partial\omega}\phi(\omega)$ where $\Psi$ is the digamma function. The term $\sigma$ denotes the renormalization of the energy level due to quantum fluctuations, $\epsilon\rightarrow\epsilon+\sigma$. The function $\sigma_\Gamma$ on the other hand is related to the $\Gamma$-renormalization. The prefactor depending on the Fermi functions relates to the fact that $\sigma_\Gamma$ contains the weighted $\Gamma$-renormalizations at different resonances, occurring with opposite signs. For details see Ref.~\onlinecite{Splettstoesser12}.
Also, see Refs.~\onlinecite{Konig99} and~\onlinecite{Splettstoesser12} for the remaining analytic expressions for the cotunneling contributions to the conductance.  

\begin{figure}
\includegraphics[scale=1]{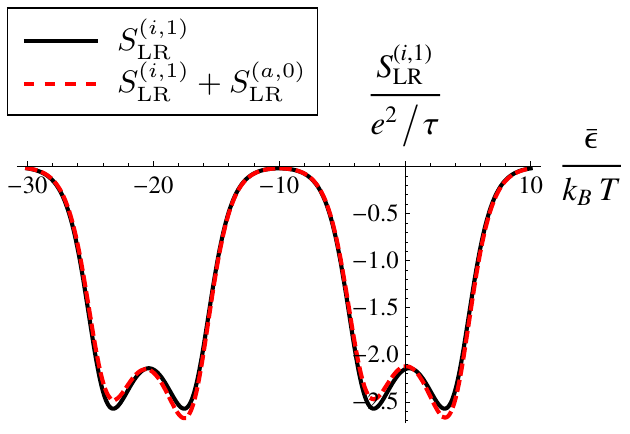}
\caption{The instantaneous noise (black solid) and the sum of instantaneous and adiabatic (red dashed) as a function of the mean energy level $\overline{\epsilon}$. The splitting of the resonances is visible for $\delta\epsilon>k_\mathrm{B}T$, here $\delta\epsilon=4k_\mathrm{B}T$ and $\delta\Gamma=0.3\Gamma$. The other parameters are $U=20k_\mathrm{B}T$, $\Omega=\Gamma/4$, and $\Gamma=0.5k_\mathrm{B}T$.}\label{fig_si_sa_shuttling}
\end{figure}

Importantly, it is the \textit{time-averaged} conductance which determines the instantaneous zero-frequency noise. This becomes clearly evident  in the black (full) line of Fig.~\ref{fig_si_sa_shuttling} for the case of large amplitudes $\delta\epsilon\gg k_\mathrm{B} T$, where the time dependence of the level position can be seen from the splitting of the resonances.  This feature is due to the fact that when the amplitude of the level energy $\delta\epsilon$ is much larger than $k_\mathrm{B}T$, the quantum dot can be at resonance for a considerably large time, even though the mean energy level $\overline{\epsilon}$ is far from resonance. When the distance from $\overline{\epsilon}$ to the resonance is equal to the amplitude at harmonic driving, the dot level is at resonance at the turning point of the sinusoid, hence the dot stays in the resonant regime for a much longer time than if the mean energy were exactly on resonance. 

Moreover, for large amplitudes, the instantaneous thermal noise part gets decreased considerably, as can be seen when comparing Fig.~\ref{fig_si_sa_nobias}(a) with Fig.~\ref{fig_si_sa_shuttling}, due to the fact that the addition energies are not always close to the resonances during the whole pumping cycle. We will show in Sec.~\ref{sec_quantized_pumping}, that $S_\text{LR}^{(i)}$ can even vanish for certain pumping schemes with large modulation amplitudes.

\subsection{Adiabatic pumping with $\epsilon(t)$ and $\Gamma_\alpha(t)$}\label{sec_pumping_ad}

While an instantaneous contribution to the noise is also present in a time-independent system, an adiabatic contribution is uniquely due to a finite time-dependent modulation. Therefore the characteristics of the pumping noise contain more information about the actual pumping process than the pumped charge only, as we will show in the following.

By assuming zero bias, but an arbitrary combination of time-dependent parameters, we find for the leading order in the tunnel coupling for the first-order in $\Omega$ correction to the noise (i.e., the pumping noise)
\begin{align}\label{eq_Sa}
S_{\mathrm{LR}}^{\left(a,0\right)}&=2e^{2}\int_{0}^{\tau}\frac{dt}{\tau}\frac{\Gamma_{\mathrm{L}}\Gamma_{\mathrm{R}}}{\Gamma^{2}}\frac{d}{dt}\left\langle \Delta \charge\right\rangle _{t}^{\left(i,0\right)}\ .
\end{align}
Importantly, also for the case of time-dependent driving, it is possible to express the zero-frequency noise exclusively in terms of observables in the dot which are local in time, and their derivatives. We immediately see that, similar to the pumping current,~\cite{Splettstoesser06} the pumping noise in the sequential tunneling regime can be non-zero only if  $\epsilon$ is one of the time-dependent parameters, such that the occupation number $\langle \charge \rangle^{(i,0)}_t$ changes in time, too. For the remainder of Sec.~\ref{sec_pumping_ad}, we choose $\Gamma_\text{L}$ as the second time-dependent parameter.

Comparing Eq.~(\ref{eq_inst_noise_Gamma1}) and Eq.~(\ref{eq_Sa}),  we see clear similarities between the structure of the pumping noise and the  instantaneous noise, which exclusively stems from thermal fluctuations. However, the adiabatic noise is due to  a working pump (i.e., two independent driving parameters) and is, therefore, a non-equilibrium effect. Hence unlike in the instantaneous case, where the system is assumed to be in equilibrium for each point in time, one cannot, in general, expect the fluctuation-dissipation relation, Eq.~(\ref{eq_FDT}), to hold for the pumping noise, and indeed we show in the following that this relation breaks down.

However, only for the case of $U=0$, the equilibrium relation can be extended to first order in $\Omega$ in the following way
\begin{equation}\label{eq_FDT_noU}
S_\text{LR}^{(a)}(U=0)=-4k_\mathrm{B}T\int_0^\tau\frac{dt}{\tau}G_t^{(a)}(U=0)\ .
\end{equation}
This relation holds for arbitrary orders in $\Gamma$, provided that the temperature is still larger than the driving frequency $\Omega$. We proved it within the framework of a scattering-matrix approach (valid for $U=0$). Equation~(\ref{eq_FDT_noU}) is a consequence of the special case where the system is both non-interacting and adiabatically driven (i.e., close to equilibrium).

In the following we show that in the presence of Coulomb interaction, the relation~(\ref{eq_FDT_noU}) breaks down already in first order in $\Gamma$. For the interacting system, the pumping noise, Eq.~(\ref{eq_Sa}), can be expressed as
\begin{equation}
\begin{split}
S_{\mathrm{LR}}^{\left(a,0\right)}=-4k_\mathrm{B}T\int_0^\tau\frac{dt}{\tau}G^{(a,0)}(t)\\-2e^2\int_0^\tau\frac{dt}{\tau}\frac{\Gamma_\text{L}\Gamma_\text{R}}{\Gamma^2}\left(\lambda_c^{(1)}-\Gamma\right)\langle \charge\rangle_t^{(i,0)}\frac{d}{dt}\langle \charge\rangle_t^{(i,0)}\ .
\end{split}
\end{equation}
It can thus be separated into a part proportional to $G^{(a,0)}$, which represents an adiabatic correction of the equilibrium fluctuation dissipation theorem, and a correction term. Importantly, strong Coulomb interaction has an impact on the charge relaxation rate, $\lambda_c^{(1)}$. It therefore influences the dynamics of the quantum dot, and thus reveals the impact of the non-equilibrium due to the time-dependent driving, in form of a deviation from the equilibrium FDT. For a non-interacting system, $\lambda_c^{(1)}=\Gamma$, and therefore the correction term vanishes. Hence it is the Coulomb interaction that prevents us to write down all the noise corrections in terms of the conductance. This is a strong indication that $S_\text{LR}^{(a,0)}$ is not uniquely of thermal origin for an interacting system.

Several works dealt with a formulation of general fluctuation relations out of equilibrium.\cite{Esposito09} In a classical regime with a stationary non-equilibrium, fluctuation relations were formulated in Refs.~\onlinecite{Andrieux07,Speck07}. Also quantum systems were investigated where the impact of weak interaction on the non-markovian behaviour was studied.~\cite{Foerster08,Utsumi10} First time-dependent considerations were made in Ref.~\onlinecite{Safi11} where so far general fluctuation relations were found for the antisymmetric noise, rather than for the symmetric noise as studied here. Extensions for the latter case are envisaged for the future.

The adiabatic noise is shown in Fig.~\ref{fig_si_sa_nobias}~(b) for the case of small pumping amplitudes. We see for one that the instantaneous noise is much larger than the adiabatic one. However, an important distinguishing feature is their symmetry: whereas the instantaneous cross-correlation is always negative, its first-order in $\Omega$ correction can be both positive or negative, going along with a sign change at each resonance. Therefore, the pumping noise can lead to an enhancement as well as a reduction of the total noise. In order to understand this behavior, it is crucial to examine the properties of the charge variance $\langle\Delta\charge\rangle$, because as we see in Eq.~(\ref{eq_Sa}) the pumping noise is directly related to its time derivative. The charge variance depends on time through its strong dependence on the energy level $\epsilon$. It is finite only when the dot level is close to the Fermi energies where the dot's charge state is not well-defined. 
At resonance where the charge variance is maximal and $\frac{\partial}{\partial\epsilon}\langle\Delta\charge\rangle=0$, the pumping noise exhibits a sign change. This shows that the pumping noise is zero whenever the charge variance is insensitive to the driving with $\epsilon$.

In addition, the pumping noise has a node at the electron-hole symmetric point $\overline{\epsilon}=-U/2$ and is antisymmetric around this point (in contrast to $S_{\text{LR}}^{(i)}$, which is symmetric).
For small amplitudes, as shown in Fig.~\ref{fig_si_sa_nobias}~(b), this leads to a shift of the purely instantaneous noise contribution, which  takes place in the same direction for both the contribution at $\bar{\epsilon}\approx 0$ and at $\bar{\epsilon}+U\approx 0$. This shift direction of both contributions depends further  on the phase-difference of the driving parameters and the coupling asymmetry of the tunnel barriers.
\footnote{We would like to remark that in the limit of $U=0$ the two resonant signals of $S_{\mathrm{LR}}^{\left(a,0\right)}$ simply combine to one signal with twice the amplitude. This behaviour differs from the instantaneous noise which - as we have pointed out in Sec.~\ref{sec_pumping_inst} - gets effectively enhanced in the interacting case, due to the enhanced relaxation rate.} Finally, the total sign of the pumping noise depends on the tunnel coupling asymmetry and the pumping direction.

Considering the adiabatic correction in the regime of large amplitudes, see Fig.~\ref{fig_si_sa_shuttling}, it results in altering the heights of the two parts of the separated peaks (as a reminder, the peaks separate due to the large amplitudes), rather than in a shift. Again, which of the contributions gets increased or decreased depends on the coupling asymmetry and the direction of the pumping.\footnote{Note that for very large amplitudes one needs to choose the frequency $\Omega$ small enough to be within the adiabatic validity regime, $\Omega\delta\epsilon\ll\Gamma k_\mathrm{B} T$.}

In order to characterize the zero-frequency noise due to pumping with respect to the average current pumped through the system per period, $\bar{I}^{(a)}$,  we define an adiabatic Fano factor
\begin{equation}
F^{(a)}=\frac{S_{\text{LR}}^{(a)}}{2e\bar{I}^{(a)}}\ .
\label{eq_Fa}
\end{equation}
Since there is no instantaneous current in the absence of a bias voltage, and thus $\overline{I}^{(i)}=0$, this quantity can be interpreted as an adiabatic correction to the ordinary Fano factor. It gives information about, e.g., whether there is a time-averaged pumping current that is free of pumping noise ($F^{(a)}$ is zero), or there is pumping noise in absence of a pumped charge ($F^{(a)}$ diverges).

Note that the direction of the dc pumping current and the sign of the related pumping noise are highly sensitive to all system parameters, which is in contrast to stationary transport. Therefore our interest in the adiabatic Fano factor $F^{(a)}$ is to examine the relation of the current and the noise including also their respective sign, differing from the motivation for the original Fano factor namely to study  the noise with respect to its Schottky limit. 
\begin{figure}
\includegraphics[scale=1]{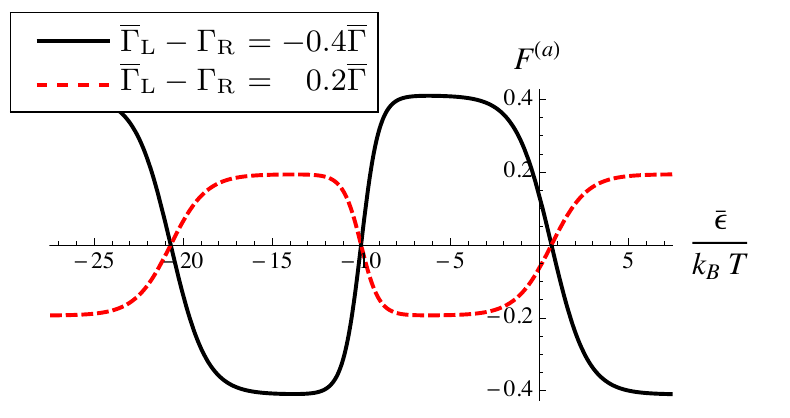}
\caption{Plot of the adiabatic Fano factor $F^{\left(a\right)}$ as a function of the mean energy level $\overline{\epsilon}$. The two curves are for $(\overline{\Gamma}_\text{L}-\Gamma_\text{R})/\overline{\Gamma}={-0.4,0.2}$ (black, red). The other parameters are $\Gamma=0.5k_\mathrm{B}T$, the pumping amplitudes $\delta\epsilon,\delta\Gamma_\text{L}=0.1\Gamma$, the phase difference between the pumping parameters $\varphi=\pi/2$, and $\Omega=0.1\Gamma$.}
\label{fig_fa_nobias}
\end{figure}

Importantly, $F^{(a)}$ is independent of the pumping direction, because both $\overline{I}^{(a)}$ and $S_{\text{LR}}^{(a)}$ change sign for a reversed pumping cycle. This means that the sign of $F^{(a)}$ can be related to intrinsic characteristics of the system (as we will see in the following).
The adiabatic Fano factor, $F^{(a)}$, is shown as a function of the average level position in Fig.~\ref{fig_fa_nobias}. We focus on the case of small amplitudes of the parameter modulation, where we can derive meaningful analytic results. In a bilinear expansion of the pumping amplitudes, we find for the adiabatic noise
\begin{equation}\label{eq_sa_bilinear}
S_\text{LR}^{(a,0)}=-2e^2\frac{\delta A}{\tau}\frac{\Gamma_\text{R}\left(\overline{\Gamma}_\text{L}-\Gamma_\text{R}\right)}{\overline{\Gamma}^3}\frac{\partial}{\partial\overline{\epsilon}}\overline{\langle\Delta \charge\rangle}_t^{(i,0)}\ ,
\end{equation}
with the area $\delta A=\int_0^\tau dt\Gamma_\text{L}\dot{\epsilon}$, which is enclosed in parameter space in one pumping cycle. As for the time-dependent system quantities, the bar over an arbitrary function, e.g. $\overline{\langle\Delta \charge\rangle}_t^{(i,0)}$, always indicates that only average parameters appear. For $\overline{\Gamma}_\text{L}=\Gamma_\text{R}$, we find that $S_\text{LR}^{(a,0)}$, in contrast to the pumped charge, is zero. Note that while the instantaneous noise and the pumped charge have even parity with respect to the particle-hole symmetric point, the pumping noise shows odd parity, $S_{\text{LR}}^{(a,0)}\left(-\overline{\epsilon}-U/2\right)=-S_{\text{LR}}^{(a,0)}\left(\overline{\epsilon}+U/2\right)$, leading to the shift of the total zero-frequency noise, discussed before. 
In lowest order in the tunnel coupling and for small amplitudes, $\delta\epsilon\ll k_\mathrm{B}T$ and $\delta\Gamma\ll \Gamma$, the adiabatic Fano factor can be approximated as
\begin{equation}\label{eq_fa_bilinear}
F^{(a)}\approx -\frac{\overline{\Gamma}_{\text{L}}-\Gamma_{\text{R}}}{\overline{\Gamma}_{\text{L}}+\Gamma_{\text{R}}}\frac{\frac{\partial^2}{\partial(\beta\overline{\epsilon})^2}\overline{\langle \charge\rangle}_t^{(i,0)}}{\frac{\partial}{\partial(\beta\overline{\epsilon})}\overline{\langle \charge\rangle}_t^{(i,0)}}\ .
\end{equation}
Evidently, the adiabatic Fano factor scales directly with the coupling asymmetry. This can be seen in Fig.~\ref{fig_fa_nobias}, where we show the Fano factor as a function of the average energy level $\overline{\epsilon}$ for different (i.e., opposite) coupling asymmetry. The latter factor can be understood as follows. If the dot level is far from resonance and the charge is exponentially suppressed, $ \overline{\langle \charge \rangle}_t^{(i,0)}\sim e^{\pm\beta\overline{\epsilon}}$, we find that the adiabatic Fano factor is constant with respect to the mean level energy $F^{(a)}=\mp\left(\overline{\Gamma}_{\text{L}}-\Gamma_{\text{R}}\right)/\overline{\Gamma}$. If the quantum dot is in resonance, i.e., the charge depends linearly on the mean level energy, $\overline{\langle \charge\rangle}_t^{(i,0)}\sim \pm\overline{\epsilon}$, we can observe a node in $F^{(a)}$. Furthermore, the Fano factor changes sign in a step-like feature at the electron-hole symmetric point, since the pumped charge and the pumped noise have different symmetries with respect to the latter. Hence, the off-resonant plateau heights are directly \textit{equal} to the coupling asymmetry, whereas the nodes occur at the resonances, and the electron-hole symmetric point. Thus,  $F^{(a)}$ displays a distinct feature for the electron-hole symmetric point, which cannot be observed by the pumping noise directly, since the noise signal itself is exponentially suppressed in this regime.
The nodes in $F^{(a)}$ correspond directly to the nodes of $S^{(a)}_\text{LR}$. If $F^{(a)}=0$ then there is no pumping noise even though there is a finite pumped charge. In the case considered here, there is no possibility to realize the opposite situation of having finite pumping noise in absence of pumped charge, in which case the adiabatic Fano factor would diverge. We will encounter this case of finite pumping noise with zero pumped charge in Sec.~\ref{sec_results_bias}.


As a next step, we study the adiabatic noise correction in next higher order in $\Gamma$, i.e., in the regime where cotunneling contributions as well as interaction-induced renormalization effects to the bare dot parameters arise due to quantum fluctuations.  It has been shown\cite{Splettstoesser06} that the average pumping current beyond sequential tunneling is due to the level renormalization term, only,
\begin{equation}\label{eq_Ia_ren}
\bar{I}^{(a,1)}=-\frac{e}{2}\int_0^\tau\frac{dt}{\tau}\frac{\Gamma_{\text{L}}-\Gamma_{\text{R}}}{\Gamma}\frac{d}{dt}\langle \charge\rangle_t^{{(i,\text{ren})}}\ .
\end{equation}
The renormalization of $\langle\charge\rangle$ is $\langle \charge\rangle_t^{(i,\text{ren})}=\sigma\frac{\partial}{\partial\epsilon}\langle \charge\rangle_t^{(i,0)}$.
Here, we present the contribution to the  zero-frequency noise in second order in the tunnel coupling. The adiabatic correction in second order is given as
\begin{equation}
S_{\text{LR}}^{(a,1)}=2e^2\int_0^\tau\frac{dt}{\tau}\frac{\Gamma_\text{L}\Gamma_\text{R}}{\Gamma^2}\frac{d}{dt}\langle\Delta \charge\rangle_t^{(i,\text{ren})}+S_{\text{LR}}^{(a,\text{corr})}\ .\label{eq_sa_second}
\end{equation}
The splitting up into two terms is motivated by Eq.~(\ref{eq_Ia_ren}). Analogously to the pumped charge, the first contribution is related to the level renormalisation, here specifically, the renormalized charge variance $\langle\Delta \charge\rangle_t^{(i,\text{ren})}=\sigma\frac{\partial}{\partial\epsilon}\langle\Delta \charge\rangle_t^{(i,0)}$. However, there is an additional  correction term
\begin{eqnarray}
S_{\text{LR}}^{(a,\text{corr})} &=&4e^2\int_0^\tau\frac{dt}{\tau}\frac{\Gamma_\text{L}\Gamma_\text{R}}{\Gamma^2}\frac{1}{\lambda_c}\frac{d}{dt}\langle \charge\rangle_t^{(i,0)}\nonumber\\ 
&& \left(\frac{2\Gamma}{U\beta}\mathcal{S}\sigma'+ (\mathcal{S}-1)\left[w-\frac{\langle\charge\rangle_t^{(i,0)}}{2}W_{\text{0d}}\right]\right.\nonumber\\
& &+ \left.(\mathcal{S}+1)\left[\widetilde{w}+\frac{\langle\charge\rangle_t^{(i,0)}}{2}W_{\text{d0}}\right]\right)\ ,\label{eq_sa_corr}
\end{eqnarray}
where we used $\sigma'=\frac{\partial}{\partial\epsilon}\sigma$ as well as $\mathcal{S}=\frac{1-f(\epsilon)-f(\epsilon+U)}{1-f(\epsilon)+f(\epsilon+U)}$. The function $\mathcal{S}$ has odd parity with respect to the electron-hole symmetric point, and gives evidence of either the electron-like or hole-like nature of the dot spectrum.\cite{Contreras12} The correction can be expressed in terms of cotunneling transition rates, for one there are rates related to transition where initial and final states are the same $w=W_{0\rightarrow\sigma\rightarrow0}=W_{\sigma\rightarrow 0\rightarrow\sigma}$, $\widetilde{w}=W_{\sigma\rightarrow \text{d}\rightarrow\sigma}=W_{\text{d}\rightarrow\sigma\rightarrow\text{d}}$, and also transitions between zero and double occupancy occur, $W_\text{0d}$ and $W_\text{d0}$. Their explicit form can be found in Ref.~\onlinecite{Splettstoesser10}. For $U=0$, we find $S_{LR}^{(a,1)}=0$, in analogy to the dc pumping current $\overline{I}^{(a,1)}$ that vanishes also. Moreover, when consulting Eq.~(\ref{eq_sa_corr}) we see that this correction term is proportional to $\langle\dot{\charge}\rangle_t^{(i,0)}$, and therefore in spite of the appearance of cotunneling terms, the adiabatic noise is still a purely resonant feature, i.e., it is exponentially suppressed far from resonance, quite like the pumped charge. 
Moreover one can show starting from Eq.~(\ref{eq_sa_second}), that in the case of symmetric coupling, $\overline{\Gamma}_\text{L}=\Gamma_\text{R}$, all adiabatic terms but $S_\text{LR}^{(a,\text{corr})}$ vanish. The behavior of this correction term is therefore quite different from the other contributions of the pumping noise which all vanish for symmetric coupling.

We consider again the adiabatic corrections of the Fano factor, see Eq.~(\ref{eq_Fa}). For a sufficiently asymmetric tunnel coupling, the only visible feature is the shift due to the level renormalization, since the correction $S_\text{LR}^{(a,\text{corr})}$ is much smaller. If we however approach the symmetric coupling case, we find that $F^{(a)}$ does not go to zero, and the only visible feature comes from $S_\text{LR}^{(a,\text{corr})}$. Importantly, $S_\text{LR}^{(a,\text{corr})}$ is even the leading contribution to the pumping noise, since also the lowest-order contribution vanishes for symmetric coupling, $S_\text{LR}^{(a,0)}$. 

Note finally, that the adiabatic correction of the Fano factor gives no account for the quality of the pump. For this purpose, one needs to include the contribution from the instantaneous, thermal noise. When choosing only $\epsilon$ and $\Gamma_\text{L}$ as pumping parameters, this is the dominant contribution, since pumping takes place close to resonance. However, when including $\Gamma_\text{R}$ as a third time-dependent parameter, it is possible to reach the quantized pumping regime. This case we will address in Sec.~\ref{sec_quantized_pumping}.

\subsection{Pumping with the barriers only}

As shown in the previous section, the pumping current and the pumping noise in lowest order in the tunneling coupling vanish if the level position is not time dependent, $\epsilon(t)=\overline{\epsilon}$. The first non-vanishing contribution of time-averaged pumping current and pumping noise for pumping with the barriers only, is due to second-order processes in the tunnel coupling. We use Eqs.~(\ref{eq_sa_second}) and (\ref{eq_sa_corr}), which are valid also for a modulation of $\Gamma_\mathrm{L}(t)$ and $\Gamma_\mathrm{R}(t)$ to evaluate the pumping noise now for pumping exclusively via the barriers. Interestingly the correction term, Eq.~(\ref{eq_sa_corr}), vanishes in this case since it requires $\langle\dot{\charge}\rangle_t^{(i,0)}\neq0$. Indeed, already for the average pumped current, when pumping with the barriers only, it has been found that pumping is uniquely due to level renormalization effects. For the only remaining  contribution to the zero-frequency pumping noise, we find
\begin{equation}
S_\text{LR}^{(a,1)}=2e^2\int_0^\tau\frac{dt}{\tau}\frac{\Gamma_\text{L}\Gamma_\text{R}}{\Gamma^2}\frac{d}{dt}\langle\Delta \charge\rangle_t^{(i,\text{ren})}\ .\label{eq_Sa_1}
\end{equation}
Note that in this situation, the pumping noise is one order of magnitude smaller than when pumping also with the level position. However, as we have pointed out already, the lowest-order contribution of the pumped charge is also zero when $\epsilon$ is constant, hence the adiabatic Fano factor is still of similar order of magnitude as when pumping with $\epsilon$ and $\Gamma_\text{L}$. Furthermore, one can show that when focussing on weak-amplitude pumping and performing an expansion in the pumping amplitudes in bilinear order, $F^{(a)}$ is equal to the expression in Eq.~(\ref{eq_fa_bilinear}). 
Therefore, in the present case of weak pumping and no bias, the shape of $F^{(a)}$ is robust with respect to the choice of any two parameters out of $\epsilon$, $\Gamma_\text{L}$ and $\Gamma_\text{R}$. Importantly, the Fano factor contains only the intrinsic properties of the system, and the pumping scheme does not obscure $F^{(a)}$. Keep in mind that this finding is true if the pumping consists of only two parameters that drive the system.

\subsection{Quantized pumping}\label{sec_quantized_pumping}

In the previous sections, we discussed predominantly the case of weak pumping, or when discussing strong pumping, we restricted ourselves to only two time-dependent parameters, with which there cannot be quantized pumping. Here, we want to consider exactly this regime where we transport one electron per cycle. This regime is of interest for a quantum standard for the current. In the regime of quantized pumping, the zero-frequency noise is strongly suppressed.~\cite{Andreev00,Makhlin01}

In order to reach a quantized pumping regime for our model of a single-level quantum dot, all three parameters $\epsilon$, $\Gamma_\text{L}$, and $\Gamma_\text{R}$ are required to be time dependent, which is in contrast to, e.g., a double quantum-dot system, where two time-dependent parameters are sufficient to obtain quantized pumping.\cite{Pothier92,Leek05} The quantized single-quantum dot pump on the other hand is most efficient when there is a phase difference of each $\pi/2$ between $\Gamma_\text{L}$ and $\epsilon$ as well as between $\epsilon$ and $\Gamma_\text{R}$, with a total phase shift of $\pi$ between $\Gamma_\text{L}$ and $\Gamma_\text{R}$.  The only possibility to reach exactly one pumped charge per cycle is when the modulation of the $\Gamma$'s is such that each $\Gamma_\alpha$ closes completely for one moment during the cycle. Thus, one achieves that the pump actually receives an electron from one lead, and half a cycle later, it can be released in the other.~\footnote{The fact that there is a phase shift of $\pi$ between the two $\Gamma$'s also assures the validity of the adiabatic condition, since the sum $\Gamma_\text{L}+\Gamma_\text{R}$ remains constant.}

\begin{figure}
\includegraphics[scale=1]{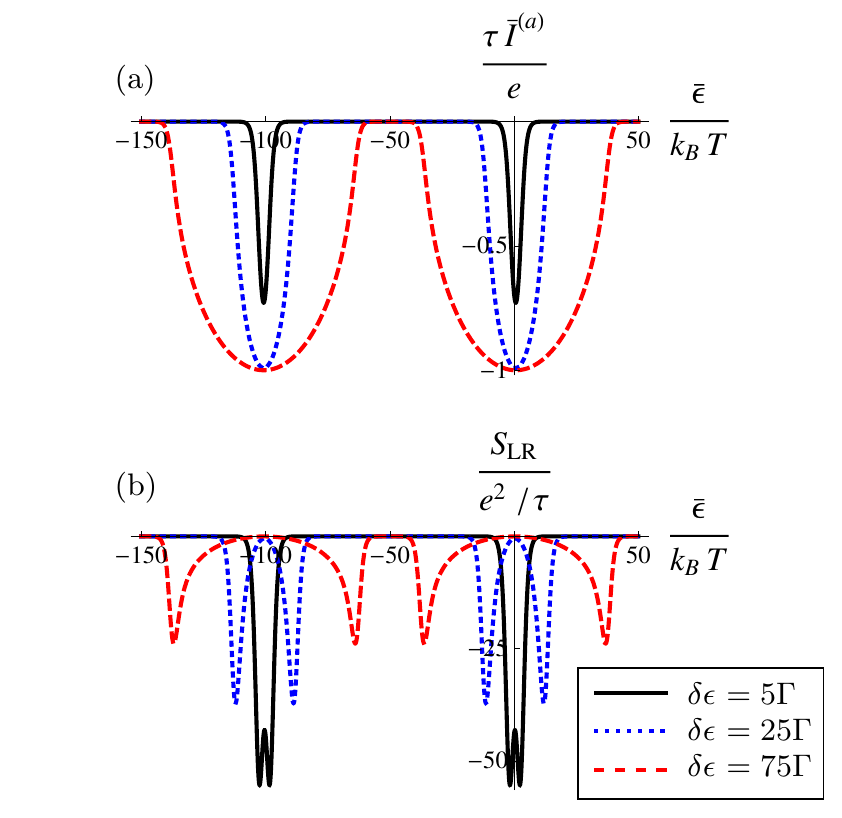}
\caption{The number of pumped charges $\tau \overline{I}_t^{(a,0)}$ (a) and the total noise cross-correlations $S_\text{LR}$ as a function of the average level energy $\overline{\epsilon}$. The time-dependent parameters are chosen as $\epsilon(t)=\overline{\epsilon}+\delta\epsilon\sin(\Omega t)$, $\Gamma_\text{L}=\Gamma/2\left[1+\sin(\Omega t-\pi/2)\right]$, and $\Gamma_\text{R}=\Gamma/2\left[1+\sin(\Omega t+\pi/2)\right]$. The amplitude of the level modulation is $\delta\epsilon=\left\{5\Gamma,25\Gamma,75\Gamma\right\}$ for the black (solid), blue (dotted), and red (dashed) line respectively. The other parameters are $U=100k_\mathrm{B} T$ and $k_\mathrm{B} T=2\Gamma$.}
\label{fig_quantised_pumping}
\end{figure}

We show the result of the number of pumped charges $\tau \overline{I}_t^{(a,0)}$ in  Fig.~\ref{fig_quantised_pumping}(a) and the corresponding total noise $S_\text{LR}=S_\text{LR}^{(i,1)}+S_\text{LR}^{(a,0)}$ in Fig.~\ref{fig_quantised_pumping}(b). We see that with increasing amplitude of the level energy $\delta\epsilon$, we reach the regime of quantized pumping, $\overline{I}=e\Omega/2\pi$. When one reaches either the resonance at the addition energy of one electron on the dot, $\overline{\epsilon}=0$, or the one to add a second electron, $\overline{\epsilon}=-U$, there is one charge transferred per cycle, for sufficiently large level energy amplitude. Considering the total noise $S_\text{LR}$ in Fig.~\ref{fig_quantised_pumping}(b) we indeed find that the noise vanishes in the regime of quantized pumping.

At the edges, i.e., when the average energy level $\overline{\epsilon}$ is $\pm\delta\epsilon$ away from the resonances, there is a remaining thermal noise contribution. Here, the dot level is brought close to resonance right at the turning point of the sinus modulation of $\epsilon$ where both tunneling barriers are open. Consequently, there are non-vanishing thermal fluctuations, and there is no perfect transmission of exactly one charge per cycle.

\section{Noise including finite bias voltage and magnetic field}\label{sec_results_bias}

We considered until now pumping at zero bias voltage. However, also the study of the current induced by a time-dependent bias opens new avenues for transport spectroscopy. In this section, we therefore generalize our previous results by allowing for a finite voltage bias, $eV(t)=\mu_\text{L}(t)-\mu_\text{R}(t)$, with the aim to  consider adiabatic pumping by the modulation of the gate and bias voltage, $\epsilon(t)=\overline{\epsilon}+\delta\epsilon\sin(\Omega t)$ and $eV(t)=e\overline{V}+e\delta V\sin(\Omega t+\varphi)$. This specific pumping prescription has a very promising application as a spectroscopy tool. It has been shown~\cite{Reckermann10,Calvo12} that when pumping with bias and gate voltages, the adiabatically pumped charge is uniquely due to a finite Coulomb interaction. If the time average of the bias is finite, the occurrence of a pumped charge reflects the sign of the coupling asymmetry to the leads, $\Gamma_\text{L}\neq\Gamma_\text{R}$; at zero average bias, the pumped charge contains information about the degeneracy of the ground state. This latter point could be proven by a study of the pumped charge in the presence of a magnetic field, which lifts the spin degeneracy of the system. 

The pumping noise turns out to have a different behavior, containing information on both the pumping and the instantaneous current. In particular, we can show that a pumping noise can persist in the absence of an adiabatically pumped charge. In this case the pumping noise indicates that the pumping current cancels in average but not at every instant of time.
In order to study the full parameter space needed for the type of spectroscopy introduced in Refs.~\onlinecite{Reckermann10,Calvo12}, we allow in the following for a finite external magnetic field as well. After deriving the general formulas for the pumping noise in Section~\ref{sec_bias_general}, we consider the specific case of time-dependent bias and gate voltage, for which the pumped charge reveals features due to Coulomb interaction in Section~\ref{sec_pump_gate_bias}.

\subsection{General expressions for current and noise}\label{sec_bias_general}

We start by adding a finite bias $V$ as well as a  Zeeman splitting $\Delta$ due to a finite magnetic field to the system and study the pumping current and the instantaneous and adiabatic zero-frequency noise for a \textit{general set} of time-dependent parameters  $\epsilon$, $\Gamma_\alpha$, $\Delta$ or $V$. As a consequence of the bias, there is also an instantaneous current, $I_t^{(i)}$ (consider Eq.~(\ref{eq_Ii}) for its evaluation), flowing in addition to the pumping current, $I_t^{(a)}$, due to time-dependent driving. Likewise, on top of thermal and pumping noise, shot noise emerges in the large-bias regime. 

The magnetic field resulting in a Zeeman splitting of the energy level adds additional structure to the system and, together with the finite Coulomb interaction, effects a coupling of the charge and spin dynamics, as will be discussed in the following.
Also, in the presence of a magnetic field a spin current is induced by a finite bias or a modulation of gates; this spin current in turn will in the following be shown to have an impact on the charge current noise.

The coupling of charge and spin dynamics is most conveniently accounted for by introducing the vector $\vec{\charge}=\left(\langle \charge\rangle,\langle s\rangle\right)^T$ as well as a current vector $\vec{I}=\left(I,I_s\right)^T$ carrying both the charge and spin on the dot and the charge and spin currents into the leads as components. The spin on the dot and the spin current are defined as the difference of the occupation or current for the two spin channels, respectively, $\left\langle s\right\rangle= \left\langle\uparrow \right\rangle-\left\langle\downarrow \right\rangle $ and $I_s=I_\uparrow-I_\downarrow$.

\textit{Pumping current.} With the abbreviations introduced above, we can give the expressions for the adiabatic charge current~\cite{Reckermann10} and the adiabatic spin current~\cite{Calvo12} in a compact form
\begin{equation}\label{eq_adiabatic_current}
\vec{I}^{(a,0)}(t)=e\mathbf{A}\frac{d}{dt}\vec{\charge}_t^{(i,0)}\ ,
\end{equation}
where the components of the instantaneous $\vec{\charge}$ are obtained through
\begin{align}
\langle \charge\rangle_t^{(i,0)}&=(-1,0,0,1)P_t^{(i,0)}\\
\langle s\rangle_t^{(i,0)}&=(0,1,-1,0)P_t^{(i,0)}\ ,
\end{align}
and $P_t^{(i,0)}$ is the stationary solution of the generalized Master equation, Eq.~(\ref{eq_master_instantaneous}). The current expression, Eq.~(\ref{eq_adiabatic_current}), shows that  time-dependent changes in spin and charge on the dot result in a current flow. The amount of the latter is found from a two-by-two response matrix $\mathbf{A}$, given in Eq.~(\ref{eq_A}), which contains information about the relaxation behavior of charge and spin in the dot. This matrix representation of the prefactor is owed to the fact that the charge and spin dynamics are coupled quantities, i.e., a change of the spin expectation value can induce a charge current and vice versa.

\begin{figure}[b]
\includegraphics[scale=0.95]{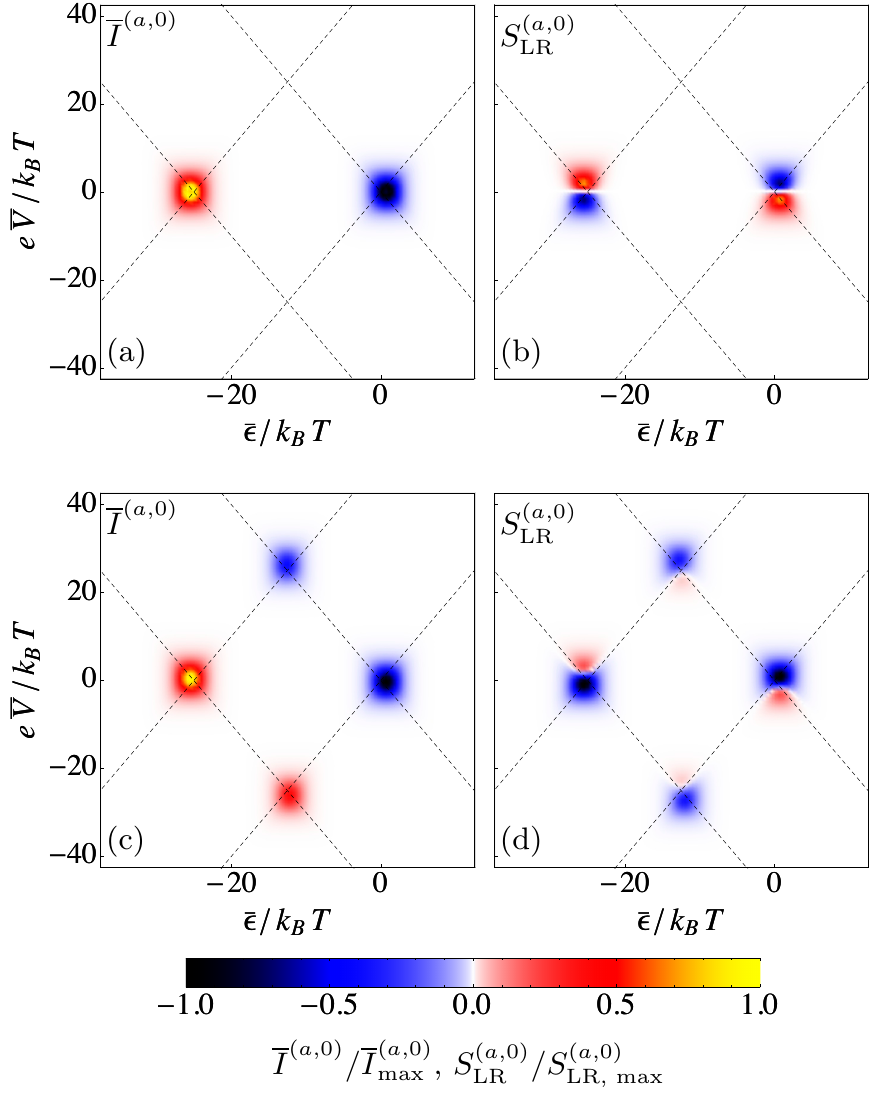}
\caption{Density plot of the pumped charge $\overline{I}^{(a,0)}$ (a) and (c), and the adiabatic noise $S_\text{LR}^{(a,0)}$ (b) and (d), as a function of the mean energy level $\overline{\epsilon}$ and the mean bias voltage $\overline{V}$, without magnetic field, $\Delta=0$. Figures (a) and (b) depict the case for symmetric tunnel coupling, $\Gamma_\mathrm{L}=\Gamma_\mathrm{R}=\Gamma/2$, and both (c) and (d) for asymmetric coupling, $\Gamma_\mathrm{L}=0.7\Gamma$ and $\Gamma_\mathrm{R}=0.3\Gamma$. The underlined dashed grid sketches the dot level resonance lines. The other parameters are $U=25k_\mathrm{B}T$, $\delta\epsilon=0.1\Gamma$, $\delta V=0.1\Gamma$, $\Omega=0.1\Gamma$, $\varphi=-\pi/2$ and $\Gamma=0.5k_\mathrm{B}T$.}
\label{fig_ia_sa_V}
\end{figure}

\begin{figure}[b]
\includegraphics[scale=0.95]{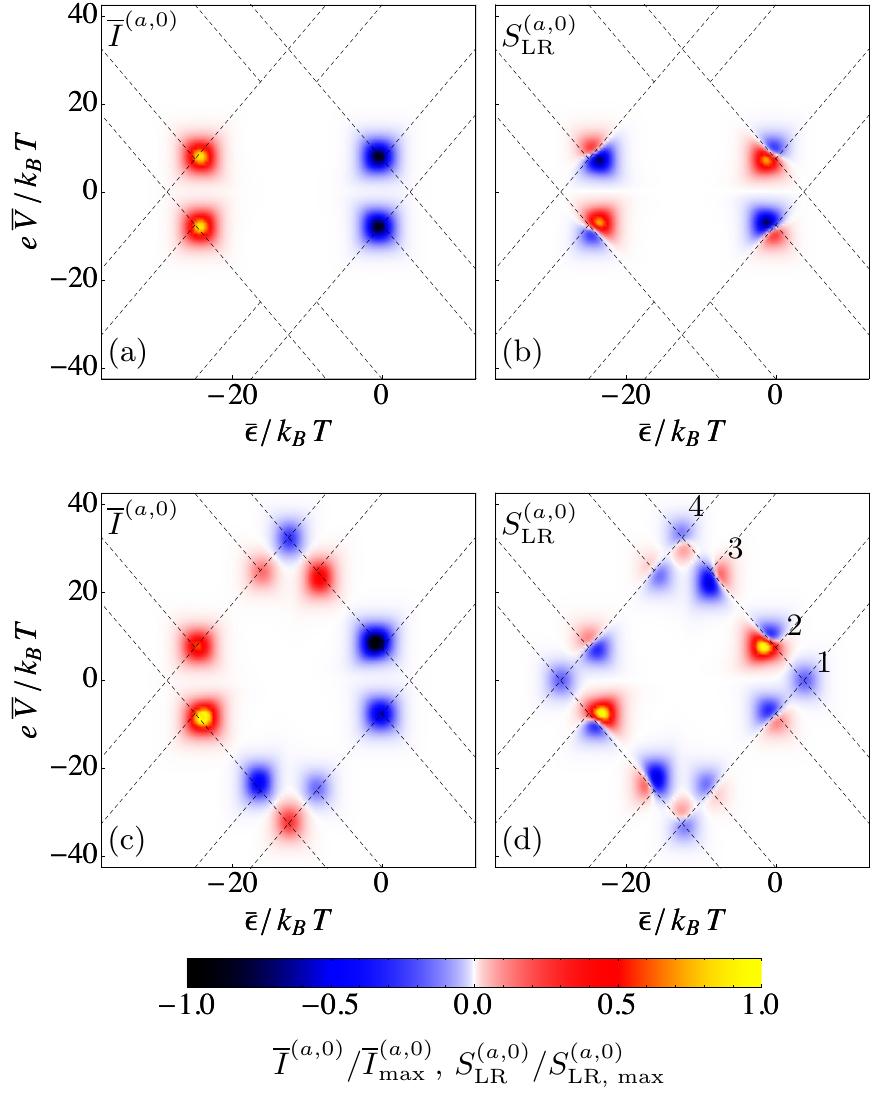}
\caption{Density plot of the pumped charge $\overline{I}^{(a,0)}$ (a) and (c), and the adiabatic noise $S_\text{LR}^{(a,0)}$ (b) and (d), as a function of the mean energy level $\overline{\epsilon}$ and the mean bias voltage $\overline{V}$, with finite magnetic field, $\Delta=7.5k_\mathrm{B}T$. Figures (a) and (b) depict the case for symmetric tunnel coupling, $\Gamma_\mathrm{L}=\Gamma_\mathrm{R}=\Gamma/2$, and both (c) and (d) for asymmetric coupling, $\Gamma_\mathrm{L}=0.7\Gamma$ and $\Gamma_\mathrm{R}=0.3\Gamma$. The underlined dashed grid sketches the dot level resonance lines. The other parameters are as in Fig.~\ref{fig_ia_sa_V}.}
\label{fig_ia_sa_B_V}
\end{figure}

For the specific case where bias and gate are chosen as pumping parameters the result for the time-averaged current is shown in Figs.~\ref{fig_ia_sa_V} and \ref{fig_ia_sa_B_V}, (a) and (c), respectively. We will discuss the shown results in detail in Sec.~\ref{sec_pump_gate_bias}.

Some features can be observed in the pumped current already in the general case. If there is either zero magnetic field or zero interaction, the dynamics of charge and spin decouple and consequently the response matrix $\mathbf{A}$ becomes diagonal. For zero magnetic field, $\Delta=0$, and arbitrary Coulomb interaction $U$ we find 
\begin{equation}
\mathbf{A}\left(\Delta=0\right)=-\frac{1}{2}\left(\begin{array}{cc}
\frac{\lambda_{c,\mathrm{L}}^{(1)}-\lambda_{c,\mathrm{R}}^{(1)}}{\lambda_{c}^{(1)}} & 0\\
0 &\frac{\lambda_{s,\mathrm{L}}^{(1)}-\lambda_{s,\mathrm{R}}^{(1)}}{\lambda_{s}^{(1)}}\end{array}\right)\ .
\end{equation}
This shows that indeed $\mathbf{A}$ contains the relaxation rates of charge and spin to the different leads $\alpha$, which, in first order in $\Gamma$ are given by
\begin{equation}
\lambda^{(1)}_{c/s,\alpha}=\Gamma_\alpha\left(1\pm f_\alpha(\epsilon)\mp f_\alpha(\epsilon+U)\right)
\end{equation}
with the Fermi function for lead $\alpha$ as $f_\alpha\left(E\right)=1/\left(1+\exp[\beta(E-\alpha eV/2)]\right)$ and the summed rates  $\lambda_{c/s}^{(1)}=\sum_\alpha\lambda_{c/s,\alpha}^{(1)}$. When the dynamics of charge and spin decouple, the pumping current is directly related to the charge relaxation rates, while the spin current is directly related to the spin relaxation rates. However for $\Delta=0$ the spin expectation value is zero at all times and so is the spin current.

Moreover, if we consider the case of non-interacting electrons,  we find, independently of the magnitude of $\Delta$, 
\begin{equation}
\mathbf{A}\left(U=0\right)=-\frac{1}{2}\frac{\Gamma_\mathrm{L}-\Gamma_\mathrm{R}}{\Gamma}\left(\begin{array}{cc}
1 & 0\\
0 & 1\end{array}\right)\ .
\end{equation}
The matrix is hence diagonal, and the charge and spin dynamics can be considered uncoupled, even with a finite magnetic field. This is because in the absence of Coulomb interaction the particles with spin $\uparrow$ and spin $\downarrow$ form two independent particle sectors. The only occurring relaxation rate for $U\rightarrow0$ is then the tunneling rate $\lambda^{(1)}_{c/s,\alpha}\rightarrow\Gamma_\alpha$. Equivalently to the two independent particle sectors one can consider any linear superposition of them, as in this case the charge $\langle \charge\rangle+1=\langle\uparrow\rangle+\langle\downarrow\rangle$ and the spin $\langle s\rangle=\langle\uparrow\rangle-\langle\downarrow\rangle$. An important consequence of the vanishing interaction is that the prefactor matrix $\mathbf{A}$ does not depend on gate and bias anymore. Therefore a \textit{time-averaged} pumped current can be induced only if at least one of the $\Gamma_\alpha$ depends on time, while the time-averaged pumped current vanishes when pumping with gate and bias. We will come back to this in Sec.~\ref{sec_pump_gate_bias}.


\textit{Zero-frequency noise.} We are interested in the properties of the noise indicating the origin of processes leading to the appearance or the suppression of a finite time-averaged pumped charge.

We start by presenting the results for the instantaneous contributions to the zero-frequency noise in the presence of a finite bias and an external magnetic field. As already stated in Sec.~\ref{sec_adiabatic}, the instantaneous noise is found to be the time integral of the expression for the stationary zero-frequency noise, where all parameters are replaced by time-dependent parameters frozen at the integration  time $t$. The stationary zero-frequency noise in the sequential tunneling regime has been calculated before in Ref.~\onlinecite{Thielmann04a}, however, analytic expressions were only provided for specific limits. Here, we find an analytic expression for the instantaneous contribution to the zero-frequency noise in a time-dependently driven system 
\begin{equation}\label{eq_instantaneous_noise}
\begin{split}
S_\text{LR}^{(i,1)}=e\int_0^\tau\frac{dt}{\tau}\left[\vec{\charge}_t^{(i,0)T}\mathbf{A}'\vec{I}_t^{(i,1)}+e\vec{a}^T\Delta\vec{\charge}_t^{(i,0)}\right]\ .
\end{split}
\end{equation}
Importantly, in contrast to the zero-bias case discussed before, see Eqs.~(\ref{eq_inst_noise_Gamma1}) and (\ref{eq_FDT}), this expression is no longer merely given by the time-averaged conductance. Now it contains time-averaged shot noise terms as well; this is clear from the first term of Eq.~(\ref{eq_instantaneous_noise}), which contains the instantaneous charge and spin current, $\vec{I}_t^{(i,1)}$, being non-zero only for a finite bias. What is more, not only is the current expression appearing in the noise, but the charge and spin densities are shown to directly couple to both the charge and spin currents via the matrix $\mathbf{A'}$. It is of very similar structure as the matrix $\mathbf{A}$ occurring in the pumping current, as can be seen in Eq.~(\ref{eq_Aprime}). 
The appearing rates are therefore the same as in the prefactor of the pumping current, see Eq.~(\ref{eq_adiabatic_current}). Here we observe that these rates reappear also in the first term of the instantaneous current-current correlation. The occurrence of rates related to charge and spin dynamics in the current noise has already been reported in a stationary system, see e.g. Refs.~\onlinecite{Davies92,Bulka99}.

The second term in Eq.~(\ref{eq_instantaneous_noise}) contains the charge and spin variance $\Delta\vec{\charge}=(\langle\Delta \charge\rangle,\langle\Delta s\rangle)^T$ and for zero bias, it reduces to the thermal noise.
We have encountered this second term, proportional to the local charge and spin fluctuations, $\Delta\vec{\charge}$, already in a simpler form in the previous case, Sec.~\ref{sec_pumping_inst}: there the instantaneous noise in the absence of a bias and a magnetic field, Eq.~(\ref{eq_inst_noise_Gamma1}), has been shown to directly depend on the charge variance and the charge relaxation rate, while in the presence of a magnetic field mixing with the spin variance occurs, which is accounted for by the prefactor vector $\vec{a}$ that contains terms related to the charge and spin relaxation see Eq.~(\ref{eq_veca}) in the Appendix. As we have stated already, for zero bias, $\vec{I}_t^{(i,1)}=0$, only the latter term survives (thermal noise). Note however that, in the large-bias limit, where shot noise effects are dominant, both terms contribute equally. Therefore, there is no clear separation in thermal and shot noise terms in this high-temperature limit with time-dependent driving, when representing the instantaneous noise according to Eq.~(\ref{eq_instantaneous_noise}). A representation for the current noise in terms of expectation values and variances of the charge and current has already been found in the time-independent regime for non-interacting electrons, zero magnetic field and large bias.\cite{Davies92}

Our main interest is here on the first-order $\Omega$ contribution to the noise for arbitrary interaction $U$ and Zeeman splitting $\Delta$. 
We here show that it can be given in terms of the dot's charge and spin and the variances thereof, as well as the charge and spin currents, in a form that reflects the structure of the instantaneous noise,
\begin{widetext}
\begin{equation}\label{eq_sa_general}
S_\text{LR}^{(a,0)}=e\int_0^\tau\frac{dt}{\tau}\left[\vec{\charge}_t^{(i,0)T}\mathbf{B}_1\vec{I}_t^{(a,0)}+\frac{d}{dt}\vec{\charge}_t^{(i,0)T}\mathbf{B}_2\vec{I}_t^{(i,1)}+\vec{\charge}_t^{(i,0)T}\mathbf{B}_3\frac{d}{dt}\vec{I}_t^{(i,1)}+\vec{\charge}_t^{(i,0)T}\mathbf{B}_4\vec{I}_t^{(i,1)}+e\,\vec{b}^T_1\frac{d}{dt}\Delta\vec{\charge}_t^{(i,0)}+e\,\vec{b}^T_2\Delta\vec{\charge}_t^{(i,0)}\right]\ .
\end{equation}
\end{widetext}
This equation represents a closed form for the pumping noise for interacting electrons and an arbitrary choice of the driving parameters.
Also in the pumping noise we find that the charge and spin expectation values appear paired with the charge and spin currents (and their time derivatives), where prefactors related to the charge and spin relaxation accompany these pairs.
The prefactors of the first four terms are collected in the matrices $\mathbf{B}_i$, given in the Appendix in Eqs.~(\ref{eq_B1}),~(\ref{eq_B2}),~(\ref{eq_B3}), and~(\ref{eq_B4}). In particular, $\mathbf{B}_4$ contains expressions proportional to the time derivatives of the pumping parameters. Like, e.g., $\mathbf{A'}$, the $\mathbf{B}_i$ have again the property that if charge and spin dynamics are independent, namely for $U=0$ or $B=0$, they become diagonal. 

Finally, we find the two latter terms in the pumping noise which
contain charge and spin variances and their time derivatives. For the general case of an arbitrary magnetic field, both the charge and spin variance influence the pumping noise, as is expressed via the prefactor vectors $\vec{b}_i$ related to the charge and spin relaxation, see Eq.~(\ref{eq_b1}) in the Appendix. Here, $\vec{b}_2$, like $\mathbf{B}_4$, contains factors proportional to the time derivatives of the pumping parameters.

In analogy to the instantaneous noise, for a vanishing magnetic field, $\Delta=0$, when spin and charge are independent, the full pumping noise can be expressed  in terms of quantities related to the charge alone, namely the charge on the dot, $\langle \charge\rangle_t^{(i,0)}$, and its variance, $\langle\Delta \charge\rangle_t^{(i,0)}$, the charge current, $I_t^{(i,1)}$, the charge relaxation rate, $\lambda_{c,\alpha}^{(1)}$, and the time derivatives of these charge-related functions.

\subsection{Adiabatic pumping with $\epsilon(t)$ and $V(t)$}\label{sec_pump_gate_bias}

We will in the following use the general results obtained above in order to study the pumping current and noise in the case where $\epsilon(t)$ and $V(t)$ are the pumping parameters. 
We divide the following discussion into two parts. First, we recapitulate the discussion of the average pumping current, $\overline{I}^{(a)}$, that has been investigated in Refs.~\onlinecite{Reckermann10,Calvo12}. In the second part, we discuss our new results of the pumping noise $S_\text{LR}^{(a)}$, and put it into relation with the pumping current $I_t^{(a)}$ and the instantaneous current $I_t^{(i)}$.

\subsubsection{Pumping current}
The results for the dc pumping current $\overline{I}^{(a,0)}$ for a finite interaction $U=25k_\mathrm{B}T$ are depicted for different parameter sets of coupling asymmetry and magnetic field in Figs.~\ref{fig_ia_sa_V} and ~\ref{fig_ia_sa_B_V}, (a) and (c). The figures display the case of weak pumping amplitudes, i.e., the modulation amplitudes of gate and bias are much smaller than $k_\text{B}T$.

The dashed lines in these figures represent the situation where either the $|0\rangle\rightarrow|\sigma\rangle$ or $|\sigma\rangle\rightarrow|2\rangle$ transition is in resonance with one of the electrochemical potentials of either the left or right lead. Due to Coulomb blockade the charge of the dot is fixed inside the diamonds formed by the dashed lines.

The first observation for the pumped charge is that there can be only a non-vanishing signal when two resonance lines cross (whenever two dashed lines meet). The reason for this is that only there, the system is  effectively sensitive to two independent pumping parameters, namely to the distance between the level position and one of the electrochemical potentials, $\epsilon(t)\pm \frac{e}{2}V(t)$. In all other regions the system is sensitive to only one of these energy differences (namely on the dashed lines in between such two points) or  not sensitive to the parameter variation at all (in the regions away from the dashed lines). In these latter cases the averaged pumped charge is identical to zero. Moreover, the dc pumping current is zero for a non-interacting quantum dot, $U=0$, because in that case the charge and spin relaxation rates become constant, and consequently there is no difference in the loading and unloading process of the pumping current (as already stated in the previous Sec.~\ref{sec_bias_general}).

What is of interest are the features of the adiabatically pumped current at the crossing points, which depend on the magnetic field and the coupling asymmetries. Figure~\ref{fig_ia_sa_V} (a) shows the pumped charge in the absence of a magnetic field, $\Delta=0$, and for symmetric coupling to the leads, $\Gamma_\mathrm{L}=\Gamma_\mathrm{R}$. The two peaks have opposite sign, which was shown to be related to the difference in the degeneracy of neighbouring charge states: for the parameter cycle chosen here, the dc pumping current is negative, at the points where the degeneracy of the ground state is increased by increasing the charge and the pumped charge is positive at the points where the degeneracy decreases by increasing the charge.

Figure ~\ref{fig_ia_sa_V} (c) shows that a finite pumped charge is observed also at the line crossings at high bias if an additional coupling asymmetry to the left and the right lead results in an asymmetry between the loading and unloading process along the pumping cycle, $\Gamma_\mathrm{L}=0.7\Gamma$.

Finally,  Fig.~\ref{fig_ia_sa_B_V} (a) and (c) depict the results for the adiabatically pumped current in the presence of a finite magnetic field, $\Delta=7.5k_\mathrm{B}T$, for symmetric and asymmetric coupling to the leads. First of all, new features are observed stemming from the spin-splitting of the energy level. Of these additional features, the signals at points $3$ and $4$, as indicated in Fig.~\ref{fig_ia_sa_B_V} (d), are visible only when the tunnel coupling is asymmetric, $\Gamma_\text{L}\neq\Gamma_\text{R}$.

Strikingly, the signal that was observed at point (1) in the absence of a magnetic field, vanishes here. This is due to the fact that the finite magnetic field lifts the degeneracy of the singly occupied state, resulting in a constant relaxation rate and therefore in a symmetric loading and unloading in the low-bias regime.  As a consequence, the pumping current can be written as a full time derivative, and its time average vanishes. To show this effect, we approximate Eq.~(\ref{eq_adiabatic_current}) in the vicinity of point $1$ for the case that point $1$ is well separated from all other contributions (namely under the assumption that $\Delta$ and $U$ are sufficiently larger than $k_\mathrm{B}T$) and find 
\begin{equation}\label{eq_Ia_point_1}
\vec{I}^{(a,0)}_t\approx-\frac{e}{2}\frac{\Gamma_\text{L}-\Gamma_\text{R}}{\Gamma}\frac{d}{dt}\vec{\charge}_t^{(i,0)}\ .
\end{equation}
Since the $\Gamma_\alpha$ are constant in time, the dc component of $\vec{I}^{(a,0)}$ is zero. Note however, that the prefactor directly shows that for symmetric coupling the symmetric contribution to the time-resolved pumping current, $\vec{I}^{(a,0)}_t=\left(\vec{I}^{(a,0)}_{\mathrm{L},t}-\vec{I}^{(a,0)}_{\mathrm{R},t}\right)/2$, at point $1$ is zero for all times, whereas for asymmetric coupling it cancels only in the time average. 
This vanishing of the pumped charge due to the lifting of the degeneracy of the ground state is important as a spectroscopy tool.\cite{Reckermann10,Calvo12}
However, from the charge current alone it is impossible to distinguish whether the pumping current vanishes on average or for all times $t$ in the low bias regime. Importantly, as we will show later, the pumping noise enables us to differentiate between exactly these two cases.


\subsubsection{Pumping noise} 

The result for the pumping noise is shown in Figs.~\ref{fig_ia_sa_V} and~\ref{fig_ia_sa_B_V}, (b) and (d), with the same set of parameters used for the discussion of  the pumping current. In general we see that also the pumping noise occurs in the vicinity of  the dashed line crossings only, since as explained above, these are the only points where the system is sensitive to two independent pumping parameters.
In contrast to the pumped charge, the features of the pumping noise display a sign change, whenever the pumped charge is finite. 

Furthermore, the pumped charge and pumping noise generally have the following different symmetry properties with respect to the average gate and bias voltage
\begin{eqnarray}\label{eq_symm_Ia}
\overline{I}^{(a,0)}(U/2-\overline{\epsilon},-\overline{V})&=&-\overline{I}^{(a,0)}(\overline{\epsilon},\overline{V})\\\label{eq_symm_Sa}
S_\text{LR}^{(a,0)}(U/2-\overline{\epsilon},-\overline{V})&=&+S_\text{LR}^{(a,0)}(\overline{\epsilon},\overline{V})\ ,
\end{eqnarray}
that is, $\overline{I}^{(a,0)}$ is antisymmetric and $S_\text{LR}^{(a,0)}$ is symmetric with respect to the point reflection at the electron-hole symmetric point, $(\overline{\epsilon},\overline{V})=(-U/2,0)$.
These relations can be proven at the level of Eqs.~(\ref{eq_adiabatic_current}) and~(\ref{eq_sa_general}). \footnote{The inversion in the point $(-U/2,0)$ results in the transformations $g_{\alpha\sigma}\rightarrow g_{\alpha\sigma}$, $\vec{\charge}_t^{(i,0)}\rightarrow-\vec{\charge}_t^{(i,0)}$, $\Delta\vec{\charge}_t^{(i,0)}\rightarrow\Delta\vec{\charge}_t^{(i,0)}$ as well as $\vec{I}_t^{(i,1)}\rightarrow-\vec{I}_t^{(i,1)}$, hence the entire noise expression remains the same and the current changes sign. Additionally, since a trajectory in the parameter space of $\epsilon(t)$, $V(t)$ keeps its orientation under this transformation, the time integral does not change the symmetry.}
The same symmetry relations as in Eqs.~(\ref{eq_symm_Ia}) and~(\ref{eq_symm_Sa}) hold also for instantaneous contributions of current, $\overline{I}_t^{(i,1)}$ and noise, $\overline{S}_\text{LR}^{(i,1)}$, respectively.

The complex structure of the pumping noise can be more easily accessed when concentrating on working points close to the zero-bias region.
We start by considering the most simple case of symmetric tunnel coupling $\Gamma_\text{L}=\Gamma_\text{R}$ and zero magnetic field, $\Delta=0$, depicted in Fig.~\ref{fig_ia_sa_V} (b). In the low bias regime the pumping noise can be approximated as
\begin{equation}\label{eq_Sa_approx}
S_\text{LR}^{(a,0)}\approx\frac{8}{\Gamma}\int_0^\tau\frac{dt}{\tau}I_t^{(i,1)}I_t^{(a,0)}\ ,
\end{equation}
provided that $U\gg k_\text{B}T$. This means that in this limiting case the pumping noise can be fully related to the product of the time-resolved currents which are of relevance here, namely the pumping current \textit{and} the instantaneous current. This expression containing current expectation values, which are again local in time, reflects the structure of the current-current correlations in a very simple product form. The noise thus contains information about the relative behaviour of these two time-resolved currents.

Remarkably, also for asymmetric tunnel-coupling $\Gamma_\text{L}\neq\Gamma_\text{R}$,  this simple formula for the low-bias regime still yields a  good qualitative approximation to the pumping noise. Therefore, we will now make use of Eq.~(\ref{eq_Sa_approx}) to understand the shape of the noise signal in detail.  For this purpose we furthermore write the pumping current in terms of a response function, $\Lambda$, as introduced in Ref.~\onlinecite{Calvo12},
\begin{equation}
I^{(a,0)} = e\Lambda\left(\dot{\vec{X}}^{\text{T}}\vec{\nabla}\langle\charge\rangle^{(i,0)}\right)\ ,
\end{equation} 
where we used a vector notation for the two pumping parameters $X^\text{T}=\left(\epsilon(t),eV(t)\right)$ and the gradient with respect to said parameters $\vec{\nabla}^\text{T}=\left(\partial_\epsilon,\frac{1}{e}\partial_V\right)$. The response function for $\Delta=0$ is given by $\Lambda=-\left(\lambda^{(1)}_{c,\text{L}}-\lambda^{(1)}_{c,\text{R}}\right)/2\lambda^{(1)}_c$.

Now, we explicitly insert the pumping current contribution into Eq.~(\ref{eq_Sa_approx}), where we expand in small pumping amplitudes, 
\begin{equation}\label{eq_S_response}
S_\text{LR}^{(a,0)}\approx\frac{8}{\Gamma}\overline{I}^{(i,1)}\overline{I}^{(a,0)}+\overline{\Lambda}\delta A\frac{8e}{\Gamma}\left(\vec{\nabla}I^{(i,1)}\times\vec{\nabla}\langle\charge\rangle^{(i,0)}\right)_3\ . 
\end{equation}
The area of one pumping cycle in parameter space is $\tau\delta A=e\int_0^\tau dt\epsilon\dot{V}$. We insert a third component equal to zero in $\vec{\nabla}$ for a well-defined cross product, where the third component of the latter appears here. 

The first term in this equation is simply a product of time-averaged currents. Since the instantaneous current always changes sign when reversing the bias and the pumping current does not change its sign within any resonance, we can generally state that this first contribution has opposite signs for opposite bias with a node at zero bias.

The different shape of the noise signals for different tunnel couplings to the leads, see Fig.~\ref{fig_ia_sa_V} (b) and (d), thus originates from the behaviour of the second contribution. Since the gradient of the current, $\vec{\nabla}I^{(i,1)}$, points mainly along the bias axis and the gradient of the charge, $\vec{\nabla}\langle\charge\rangle^{(i,0)}$, is mainly directed along changes in the gate, the cross product has a constant sign for all values of $\overline{\epsilon}$ and $\overline{V}$. Hence, for a fixed pumping cycle, only the averaged response function $\overline{\Lambda}$ is strongly sensitive to the coupling asymmetry. It therefore governs the symmetry behavior of the pumping noise at low bias. 

For symmetric coupling the response function $\overline{\Lambda}$ has a node at zero bias, making the second contribution of Eq.~(\ref{eq_S_response}) of equal shape as the first. This explains the antisymmetric behavior (with respect to $V$) of the pumping noise at low bias for symmetric tunnel coupling, shown in Fig.~\ref{fig_ia_sa_V}~(b). However, when the tunnel coupling is strongly asymmetric, as shown in Fig.~\ref{fig_ia_sa_V}(c) and (d) for $\Gamma_\mathrm{L}=0.7\Gamma$ and $\Gamma_\mathrm{R}=0.3\Gamma$, the response function $\overline{\Lambda}$ does not exhibit any sign change. Therefore the antisymmetric behavior of the noise is lifted, see Fig.~\ref{fig_ia_sa_V}~(d).

Concluding the discussion of the $\Delta=0$ case, we consider the pumping noise signal in the high bias regime. Also at high bias, pumping noise occurs whenever the pumped charge is finite, as can be seen in Fig.~\ref{fig_ia_sa_V}~(d). The approximate Eq.~(\ref{eq_Sa_approx}) is no longer valid in this regime, and in order to understand the detailed behavior, the full Eq.~(\ref{eq_sa_general}) has to be considered. Nonetheless, this pumping noise exhibits a similar sign change as in the low-bias regime, the orientation of which still reflects the respective sign of the time-averaged pumping current and the instantaneous current (we will describe this effect in more detail in the following $\Delta\neq0$ case).

Strikingly, when considering a finite magnetic field, the pumping noise can persist even when the pumped charge vanishes. In this particular case the pumping noise does not exhibit a sign change. This effect can be observed in the case of asymmetric coupling in the low bias regime, see Figs.~\ref{fig_ia_sa_B_V} (c) and (d). In order to reproduce this remarkable feature, we extend our discussion of the pumping noise in the low-bias regime to the case of a  finite magnetic field, $\Delta\neq0$. Importantly,  Eq.~(\ref{eq_Sa_approx}) is a valid quantitative approximation  for finite $\Delta$ independently of the tunnel coupling asymmetry,  given that $U,\Delta\gg k_\text{B}T$. 

Since we are here interested in a situation in which the average pumped current vanishes, the first term of Eq.~(\ref{eq_S_response}) never contributes and the full behavior can be understood from the time-averaged response function. In the limit $U,\Delta\gg k_\text{B}T$ and for low bias it is given by
\begin{equation}
\bar{\Lambda}(\Delta) = -\frac{\Gamma_\mathrm{L}-\Gamma_\mathrm{R}}{2\Gamma}\ ,
\end{equation}
see also Eq.~(\ref{eq_Ia_point_1}). This response function is independent of gate and bias, which explains the missing sign change. Furthermore,  $\bar{\Lambda}(\Delta)$ takes a constant value different from zero for a finite coupling asymmetry.  For $\Gamma_\text{L}\neq\Gamma_\text{R}$, the time-resolved pumped current is different from zero, even though its average vanishes. This explains the finite contribution to Eq.~(\ref{eq_Sa_approx}).

However, when the coupling is symmetric, the response function $\overline{\Lambda}(\Delta)$  is exactly zero and so is the pumping current at every instant of time. Therefore also the pumping noise vanishes, see Fig.~\ref{fig_ia_sa_B_V}~(b).
This shows that with the help of  the pumping noise - which, as we want to stress, is itself a time averaged quantity - one can distinguish at low bias whether the pumping current is zero at all times or whether it is only its time average which vanishes.

Finally, we also remark on the high-bias pumping noise occurring in the presence of  a finite magnetic field, as shown in Figs.~\ref{fig_ia_sa_B_V} (b) and (d).
The signals, such as points 2 to 4,  go along with a finite pumped charge and hence exhibit the sign change observed and discussed before. As mentioned already for the high bias noise signals in the case where $\Delta=0$, 
the orientation of the nodes in the pumping noise reflect the respective directions of instantaneous current and pumped charge. Namely, when going away from the electron-hole symmetric point, the pumping noise changes from positive to negative (negative to positive) when pump and bias work in the opposite (same) direction. This fact also manifests in the symmetry relations, Eqs.~(\ref{eq_symm_Ia}) and~(\ref{eq_symm_Sa}), derived earlier.

\section{Conclusion}\label{sec_conclusion}

In this paper, we developed a formalism for the calculation of the zero-frequency pumping noise in the adiabatic driving regime in the presence of strong Coulomb interaction and a non-equilibrium due to an arbitrary externally applied bias voltage. We found analytic expressions for the pumping noise containing expectation values of the dot occupation and the current, which are \textit{local in time}. This allows us  to individuate contributions originating from thermal noise, shot noise and pumping noise.

In a first step we applied the developed formalism to the case of pure pumping, in absence of an external bias voltage, up to second order in the tunnel coupling $\Gamma$. In zeroth order in the driving frequency, the zero-frequency noise fulfills a time-averaged version of the fluctuation-dissipation theorem. We find that for the correction in first order in the driving frequency, the fluctuation-dissipation theorem breaks down, \textit{uniquely due to the nonvanishing Coulomb interaction}. This is true already in first order in the tunnel coupling $\Gamma$.
We study the characteristic properties of the pumping noise based on an analysis of the adiabatic correction of the Fano factor. We find that it exhibits information about the coupling asymmetry to the leads and has a distinct feature, i.e., a step with a sign change, at the electron-hole symmetric point. Interestingly, the adiabatic correction of the Fano factor is insensitive to the specific choice of pumping parameters.

In the second part of this manuscript we addressed the pumping noise in the presence of a finite - possibly large - bias, and eventually including a magnetic field. We computed an explicit analytical expression for the pumping noise valid for arbitrary bias and magnetic field strength, as well as an arbitrary choice of time-dependent parameters. Based on these we were able to show how the charge dynamics, and (in the case of a finite magnetic field) their interplay with spin dynamics, appear in the noise. In the specific case of pumping with gate and bias voltage as time-dependent parameters in particular in presence of a magnetic field, we find that there can be pumping noise in the absence of pumped charge. The appearance of this additional noise signal can be used to identify whether the time-resolved pumping current vanishes or whether it averages out after one period. More generally the pumping noise reveals the respective direction of pumping current and the current induced by a  bias.

\acknowledgments
We acknowledge fruitful discussion with Hern\'{a}n Calvo, Federica Haupt, Oleksiy Kashuba, and Michael Moskalets. Financial support was provided by the Ministry of Innovation, NRW.

\appendix

\section{Adiabatic expansion in Laplace space}\label{appendix_adiabatic_laplace}
In this section we want to introduce a general scheme for an adiabatic expansion, as derived  in Ref.~\onlinecite{Kashuba12}. In a general case of two functions that depend on two times, $A\left(t,t'\right)$ and $B\left(t,t'\right)$ we can write the Laplace transform of their convolution $(A\circ B)(t,t')=\int_{t'}^tdt_1A(t,t_1)B(t_1,t')$ as
\begin{align}
\left(A\circ B\right)\left(t,z\right)&=\int_{-\infty}^{t}dt'e^{-z\left(t-t'\right)}\int_{t'}^{t}dt_{1}A\left(t,t_{1}\right)B\left(t_{1},t'\right)\nonumber\\
&=
\int_{-\infty}^{t}dt_{1}e^{-z\left(t-t_1\right)}A\left(t,t_{1}\right)B\left(t_{1},z\right)
\nonumber\\
&=e^{\partial_{z}^{A}\partial_{t}^{B}}A\left(t,z\right)B\left(t,z\right)\label{eq_AB_laplace}
\end{align}
where, as before, the Laplace transform is defined as $A(t,z)=\int_{-\infty}^tdt'e^{z(t'-t)}A(t,t')$ and analogously for $B$. The operator $\partial_{z}^{A}$ is the derivative with respect to $z$ acting on $A$ and $\partial_{t}^{B}$ the time derivative of $B$, respectively. From the first to second line in Eq.~(\ref{eq_AB_laplace}) we swapped the integral limits, and from the second to third line, we performed a Taylor expansion of $B$ around time $t$, i.e., $B\left(t_1,z\right)=\sum_{n=0}^\infty\frac{1}{n!}\left(t_1-t\right)^n\partial^n_t B\left(t,z\right)$.

The adiabatic approximation takes into account only terms up to first order in $\Omega$, ergo only first-order derivatives of $\partial_{t}$ are considered. Thus, as long as $\partial_zA(t,z)\partial_tB(t,z)$ is sufficiently small\cite{Kashuba12}
\begin{equation}
\left(A\circ B\right)\left(t,z\right)\approx\left(1+\partial_{z}^{A}\partial_{t}^{B}\right)A\left(t,z\right)B\left(t,z\right).
\end{equation}
If each of the objects has an adiabatic expansion of its own $A\left(t,z\right)\approx A^{\left(i\right)}\left(t,z\right)+A^{\left(a\right)}\left(t,z\right)$, a consistent expansion of the convolution can be given as follows. The instantaneous part (zeroth order in $\Omega$) is
\begin{equation}
\left\{ \left(A\circ B\right)\left(t,z\right)\right\} ^{\left(i\right)}= A^{\left(i\right)}\left(t,z\right)B^{\left(i\right)}\left(t,z\right)
\end{equation}
and the adiabatic correction (first order in $\Omega$) can be evaluated as
\begin{equation}
\begin{split}
\left\{ \left(A\circ B\right)\left(t,z\right)\right\} ^{\left(a\right)}= A^{\left(i\right)}\left(t,z\right)B^{\left(a\right)}\left(t,z\right)\\+A^{\left(a\right)}\left(t,z\right)B^{\left(i\right)}\left(t,z\right)+\partial_{z}A^{\left(i\right)}\left(t,z\right)\dot{B}^{\left(i\right)}\left(t,z\right).
\end{split}\label{eq_adiabatic_exp_general}
\end{equation}
Eventually, for all expressions we encounter, the relevant limit is $z\rightarrow0^+$. Nonetheless, we keep $z$ finite at this stage, because we will in some cases need to deal with derivatives with respect to the Laplace variable $z$.
\subsubsection*{Kinetic equation}
In Ref.~\onlinecite{Kashuba12} this way of writing a general convolution in time space was also used for a handy formulation of the adiabatic expansion of the kinetic equation. In this case the starting point is the equation
\begin{eqnarray}
\frac{dP(t)}{dt} &  = & \int_{-\infty}^{t}dt' W(t,t')P(t')\\
& = &  \int_{-\infty}^{t}dt' W(t,t')\sum_{n=0}^{\infty} \frac{1}{n!}(t'-t)^n\frac{d^n}{dt^n}P(t)\nonumber
\end{eqnarray}
This expansion is readily identified as the formal series
\begin{equation}\label{eq_kineq_general}
\frac{dP(t)}{dt} =  \left.e^{\partial_z^W\partial_t^P}W_t(z)P(t)\right|_{z=0}
\end{equation}
and can serve as the starting point for a frequency expansion of the kinetic equation in arbitrary order. This equation is similar to Eq.~(\ref{eq_AB_laplace}) except for the fact that $P(t')$ is a function of a single time and, therefore, its Laplace transform does not occur in the final result, Eq.~(\ref{eq_kineq_general}). As pointed out in Ref.~\onlinecite{Kashuba12}, this can however be treated in complete analogy, when artificially introducing the two-time function $P(t_1,t')=P(t_1)\delta(t_1-t'-0^+)$ and writing the kinetic equation as
\begin{eqnarray}
\frac{dP(t)}{dt} &=& \lim_{z\rightarrow0^+}(W\circ P)(t,z)\\\nonumber
&=& \lim_{z\rightarrow0^+}\int_{-\infty}^{t}dt'e^{-z\left(t-t'\right)}\int_{t'}^{t}dt_{1}W(t,t_1)\\\nonumber
&& \times P(t_1)\delta(t_1-t'-0^+)\\
&=& \int_{-\infty}^{t}dt' W(t,t')P(t')
\end{eqnarray}
Here the Laplace transform of $P(t_1,t')$ is easily found to be
\begin{eqnarray}
P(t,z)  & = & \int_{-\infty}^{t}dt'e^{-z\left(t-t'\right)}P(t)\delta(t-t'-0^+)\nonumber\\
 & = & P(t)
\end{eqnarray}
In the following we treat functions of a single time when occurring in the previously introduced brackets in this way, such that
\begin{eqnarray}
\left\{ \left(W\circ P\right)\left(t,z\right)\right\} ^{\left(i\right)} & = & W_t^{\left(i\right)}\left(z\right)P_t^{\left(i\right)}\\
\left\{ \left(W\circ P\right)\left(t,z\right)\right\} ^{\left(a\right)} & = &  W_t^{\left(i\right)}\left(z\right)P_t^{\left(a\right)}\\
&+&W_t^{\left(a\right)}\left(z\right)P_t^{\left(i\right)}
+\partial_{z}W_t^{\left(i\right)}\left(z\right)\dot{P}_t^{\left(i\right)}.\nonumber
\end{eqnarray}

\section{Adiabatic expansion of the zero-frequency noise}\label{appendix_noise_adiabatic}
Here we want to evaluate the adiabatic correction to all terms that contribute to the noise. The instantaneous contribution corresponds to the results obtained in Ref.~\onlinecite{Thielmann04a,Thielmann05}, where in our case all parameters depend parametrically on time. Here we concentrate on the contributions due to the adiabatic corrections. In order to be able to refer to the noise contributions in Eq.~(\ref{eq_noisepart_det}) in a simple fashion let us name them
\begin{widetext}
\begin{align}
S_{\alpha\beta}^{A}\left(t\right)&=e^{2}\text{e}^\text{T}\int_{t_{0}}^{t}dt_{1}W_{I_{\alpha}I_{\beta}}\left(t,t_{1}\right)P\left(t_{1}\right)
+\left(\alpha\leftrightarrow\beta\right)\\
S_{\alpha\beta}^{B}\left(t\right)&=e^{2}\text{e}^\text{T}\int_{t_{0}}^{t}dt_{1}\int_{t_{0}}^{t_{1}}dt_{2}\int_{t_{0}}^{t_{2}}dt_{3}W_{I_{\alpha}}\left(t,t_{1}\right)\Pi\left(t_{1},t_{2}\right)W_{I_{\beta}}\left(t_{2},t_{3}\right)P\left(t_{3}\right)\
+\left(\alpha\leftrightarrow\beta\right)\\
S_{\alpha\beta}^{C_{1}}\left(t\right)&=-e^{2}\text{e}^\text{T}\int_{t_{0}}^{t}dt_{1}\int_{t_{0}}^{t_{1}}dt_{2}\int_{t_{0}}^{t_{2}}dt_{3}W_{I_{\alpha}}\left(t,t_{1}\right)P\left(t_{1}\right)\otimes \text{e}^\text{T}W_{I_{\beta}}\left(t_{2},t_{3}\right)P\left(t_{3}\right)
+\left(\alpha\leftrightarrow\beta\right)\\
S_{\alpha\beta}^{C_{2}}\left(t\right)&=-e^{2}\text{e}^\text{T}\int_{t_{0}}^{t}dt_{1}\int_{t_{1}}^{t}dt_{2}\int_{t_{0}}^{t_{2}}dt_{3}W_{I_{\alpha}}\left(t,t_{1}\right)P\left(t_{1}\right)\otimes \text{e}^\text{T}W_{I_{\beta}}\left(t_{2},t_{3}\right)P\left(t_{3}\right)+\left(\alpha\leftrightarrow\beta\right)\label{eq_app_S_timespace}
\end{align}
\end{widetext}
where both $S^{C_1}_{\alpha\beta}$ and $S^{C_2}_{\alpha\beta}$ 
stem from the term containing the product
of current expectation values. To obtain the two contributions, we simply split the middle time integral within the interval $\left[t_{0},t\right]$ into one from $t_{0}$ to $t_{1}$ and one from $t_{0}$ to $t$. This separation is used because $S_{\alpha\beta}^{C_1}$ is of the same structure as $S_{\alpha\beta}^{B}$, and can be treated on the same footing. Then we make use of the fact  that for $\left|t_1-t_{0}\right|\rightarrow\infty$, the propagator takes the form $\Pi\left(t_{1},t_{0}\right)= P\left(t_{1}\right)\otimes \text{e}^\text{T}$. This replacement we do in both $S_{\alpha\beta}^{C_{1}}\left(t\right)$ and $S_{\alpha\beta}^{C_{2}}\left(t\right)$.

The adiabatic expansion of the first term $S^{A}_{\alpha\beta}$ is done in a straightforward manner, by performing the expansion as in Eq.~(\ref{eq_adiabatic_exp_general}). Note that in the end, the adiabatic expansion is directly analogous to the one of the kinetic equation, see Eq.~(\ref{eq_master_adiabatic})
\begin{equation}
\begin{split}
\left\{ S_{\alpha\beta}^{A}\left(t\right)\right\} _{t}^{\left(a\right)} =  e^{2}\text{e}^\text{T}\left(W_{I_{\alpha}I_{\beta},t}^{\left(i\right)}P_{t}^{\left(a\right)}+W_{I_{\alpha}I_{\beta},t}^{\left(a\right)}P_{t}^{\left(i\right)}\right.\\\left.+\partial W_{I_{\alpha}I_{\beta},t}^{\left(i\right)}\frac{d}{dt}P_{t}^{\left(i\right)}\right) +\left(\alpha\leftrightarrow\beta\right).
\end{split}
\end{equation}
The term $S^{C_1}_{\alpha\beta}$ can be combined with the term $S^B_{\alpha\beta}$, introducing the object $\overline{\Pi}\left(t,t'\right)=\Pi\left(t,t'\right)-P\left(t\right)\otimes\text{e}^\text{T}$, see Eq.~(\ref{eq_dec_prop}).
The two terms that are combined here are depicted in Figs.~\ref{fig_contour}~(b) and (d) and contain all those terms in which the times at which the two currents are taken do not lie within the same non-irreducible part of the contour. We discuss the properties of the resulting object $\overline{\Pi}\left(t,t'\right)$  in more detail in Appendix \ref{appendix_decaying_propagator}. The outcome of this combination is a convolution of four objects. Thus, it can be handled analogously to the term $S^{A}_{\alpha\beta}$. We perform its adiabatic expansion as
\begin{eqnarray}
&&\left\{ S_{\alpha\beta}^{B}\left(t\right)+S_{\alpha\beta}^{C_{1}}\left(t\right)\right\} _{t}^{\left(a\right)} = e^{2}\text{e}^\text{T}\left[\left\{W_{I_{\alpha}}\overline{\Pi}\right\}_{t}^{\left(i\right)}\left\{ W_{I_{\beta}}P\right\} _{t}^{\left(a\right)}\right.\nonumber\\
&& +  \left.\left\{ W_{I_{\alpha}}\overline{\Pi}\right\} _{t}^{\left(a\right)}\left\{W_{I_{\beta}}P\right\}_{t}^{\left(i\right)}
+\partial\left\{W_{I_{\alpha}}\overline{\Pi}\right\}_{t}^{\left(i\right)}\frac{d}{dt}\left\{W_{I_{\beta}} P\right\}_{t}^{\left(i\right)}\right] \nonumber\\
&&+\left(\alpha\leftrightarrow\beta\right)
\end{eqnarray}
With the help of the rules for the evaluation of the brackets we can successively evaluate all contributing objects. 

The adiabatic expansion of the remaining term, $S_{\alpha\beta}^{C_{2}}\left(t\right)$, can be treated in an analogous way only after some transformations. By exchanging the first two of the time integrals in Eq.~(\ref{eq_app_S_timespace}) and setting $t_0\rightarrow-\infty$, we write
\begin{eqnarray}
S_{\alpha\beta}^{C_{2}}\left(t\right) & = & -\int_{-\infty}^{t}dt_2 \ e \mathrm{e}^\mathrm{T}\int_{-\infty}^{t_2}dt_1W_{I_{\alpha}}(t,t_1)P(t_1)\otimes\nonumber\\
& & e \mathrm{e}^\mathrm{T}\int_{-\infty}^{t_2}dt_3W_{I_{\beta}}(t_2,t_3)P(t_3)\ .
\end{eqnarray}
We identify in the second line of this equation the expression for the current $I_\beta(t_2)$. 
Furthermore, we define the two-time object 
\begin{eqnarray}
\tilde{I}_\alpha(t,t_2) & = &  e \mathrm{e}^\mathrm{T}\int_{-\infty}^{t_2}dt_1W_{I_{\alpha}}(t,t_1)P(t_1)\ ,
\end{eqnarray}
which is related to the current $I_\alpha(t)$ via $I_\alpha(t)=\tilde{I}_\alpha(t,t)$.
With this definition we are able to write the term $S_{\alpha\beta}^{C_{2}}\left(t\right)$ as the convolution
\begin{eqnarray}
S_{\alpha\beta}^{C_{2}}\left(t\right) & = & -\int_{-\infty}^{t}dt_2 \tilde{I}_\alpha(t,t_2) I_\beta(t_2)\\
& = & -e^{\partial_z^{\tilde{I}}\partial_t^I }\tilde{I}_\alpha(t,z)I_\beta(t)
\end{eqnarray}
The only step necessary to fully evaluate this expression is the calculation of the Laplace transform of the object $\tilde{I}_\alpha(t,t_2)$. We find
\begin{eqnarray}
\tilde{I}_\alpha(t,z) & = & \int_{-\infty}^{t}d t_2e^{-z(t-t_2)} \ e\mathrm{e}^\mathrm{T}\int_{-\infty}^{t_2}dt_1W_{I_{\alpha}}(t,t_1)P(t_1)\nonumber\\
&  = &  e\mathrm{e}^\mathrm{T}\int_{-\infty}^{t}\frac{d t_1}{z}\left[1-e^{-z(t-t_1)}\right]W_{I_{\alpha}}(t,t_1)P(t_1)\nonumber\\
& = & \frac{1}{z}\left[I_\alpha(t,z=0)-I_\alpha(t,z)\right]
\end{eqnarray}
This allows us to perform an expansion of $\tilde{I}_\alpha(t,z)$ in the Laplace frequency $z$. Of this expansion we need the contribution in zeroth and in first order in $z$ in order to evaluate the expansion of the zero-frequency noise in the driving frequency $\Omega$. We find
\begin{eqnarray}\label{eq_Itilde}
\tilde{I}_\alpha(t,z\rightarrow0^+) & = & -e\mathrm{e}^\mathrm{T}\partial W_{I_{\alpha},t}P_t\\\label{eq_dItilde}
\left.\partial_z\tilde{I}_\alpha(t,z)\right|_{z\rightarrow0^+} & = & - e\mathrm{e}^\mathrm{T}\frac{1}{2!}\partial^2 W_{I_{\alpha},t}P_t
\end{eqnarray}
The evaluation of $S_{\alpha\beta}^{C_{2}}\left(t\right)$ in zeroth and first order in the driving frequency in terms of the before introduced brackets 
\begin{eqnarray}\label{eq_Itilde_i}
\left\{S_{\alpha\beta}^{C_{2}}\left(t\right)\right\}_t^{(i)} & = & \left\{-\tilde{I}_\alpha I_\beta\right\}_t^{(i)}\\\label{eq_Itilde_a}
\left\{S_{\alpha\beta}^{C_{2}}\left(t\right)\right\}_t^{(a)} & = & \left\{-\tilde{I}_\alpha I_\beta\right\}_t^{(a)}
\end{eqnarray}
is now straightforward.

\section{Diagrammatic rules}\label{appendix_diagrammatic_rules}
In this section, we summarize the rules to diagrammatically evaluate the various contributions to the kernel and the current kernels as given in Refs.~\onlinecite{Konig96a,Thielmann04a,Splettstoesser06} and the new terms occurring in the pumping noise. The example shown in Fig.~\ref{fig_example_diagram} is a diagram that combines features of all rules listed below. The rules for the instantaneous kernel $W_{t}^{\left(i\right)}\left(z\right)$ in Laplace space are:

\begin{figure}
\includegraphics[scale=1]{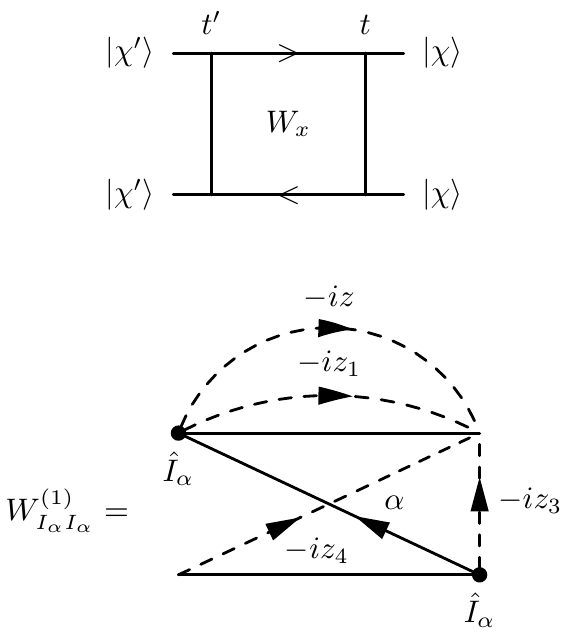}
\caption{Upper graph: general scheme of a diagram element, where $t'<t$. The index $x$ stands for the different kinds of diagrams, $W$, $W_{I_\alpha}$, and $W_{I_\alpha I_\beta}$. The forward part of the Keldysh contour is the upper line going from left to right, the lower line represents the backward part of the contour. Lower graph: example for one specific diagram first order in $\Gamma$, of type $W_{I_\alpha I_\beta}$. The solid arrow line with index $\alpha$ is representing a tunneling line where an electron is tunneling in and out of reservoir $\alpha$. The solid dots represent external vertices due to the current operator. The dashed lines represent the extra imaginary energy lines that are used for either Laplace derivatives or adiabatic corrections of diagrams.}
\label{fig_example_diagram}
\end{figure}

\begin{enumerate}
\item\label{rule1} Draw all topologically different diagrams with $n$ tunneling lines connecting pairs of vertices. Assign reservoir index $\alpha$, energy $\omega$ and spin $\sigma$ to each line. Also, assign state index $\chi$ with the corresponding energy $E_{\chi}\left(t\right)$ to each element of the Keldysh contour connecting two vertices. Additionally, draw an external line from the upper leftmost to the upper rightmost corner of the diagram, carrying an external frequency $-iz$.

\item For each time segment between two adjacent vertices (they may lie on the same or on different branches of the Keldysh contour) assign the resolvent $1/\Delta E\left(t\right)$ where $\Delta E\left(t\right)$ is the difference of all backward going minus all forward going energies.

\item 
Each vertex containing a dot operator $d_\sigma^{(\dagger)}$ gives rise to a matrix element $\left\langle \chi'\right| d_\sigma^{(\dagger)}\left|\chi\right\rangle$ where $\chi$ ($\chi'$) is the dot state entering (leaving) the vertex with respect to the Keldysh contour. For the single-level quantum dot in particular, the transitions between the doubly occupied state and the singly occupied state with an electron with spin down, $|2\rangle\rightarrow\left|\downarrow\right\rangle$ and $\left|\downarrow\right\rangle\rightarrow|2\rangle$, pick up a minus sign, $-1$, due to Pauli's principle. All other transitions do not acquire an additional factor.

\item Each tunneling line with index $\alpha$ contributes with a factor of $\frac{1}{2\pi}\Gamma_{\alpha}f_{\alpha}\left(\omega\right)$ if the line is going backward with respect to the closed time path and a factor of $\frac{1}{2\pi}\Gamma_{\alpha}f_{\alpha}\left(-\omega\right)$ if it is going forward. 

\item Each diagram takes up a prefactor of $\left(-i\right)\left(-1\right)^{b}\left(-1\right)^{c}$, where $b$ is the number of electron operators (due to internal, namely tunneling, vertices) on the backward Keldysh contour, and $c$ is the number of crossings of tunneling lines.

\item\label{rule6} Sum over all diagrams that contribute to the same kernel element.
\end{enumerate}

Next, we want to give the additional rules for the blocks containing one or two current operators, $W_{I_{\alpha},t}^{\left(i\right)}\left(z\right)$ and $W_{I_{\alpha}I_{\beta},t}^{\left(i\right)}\left(z\right)$. The diagrams contributing to $W_{I_{\alpha}}$ and $W_{I_{\alpha I_\beta}}$ can be directly derived from the rules for the kernel $W$. This direct relation arises due to the fact that the current operator, defined as $\hat{I}_\alpha=ei\sum_{k\sigma}\left[\gamma_\alpha(t)d_\sigma^\dagger c_{\alpha k\sigma}-\text{h.c.}\right]$ is of analogous structure as the tunneling part of the Hamiltonian, see Eq.~(\ref{eq_hamiltonian}), which gives rise to the tunneling vertices in $W$. The replacement of a tunnel vertex by a current operator (external vertex) results in factors $\pm1$  with respect to rule 5, depending on the position of $\hat{I}_\alpha$ on the contour. Furthermore, an additional sign occurs depending on whether a tunneling line is incoming or outgoing, due to the shape of the current operator.

\begin{figure}
\includegraphics[scale=0.9]{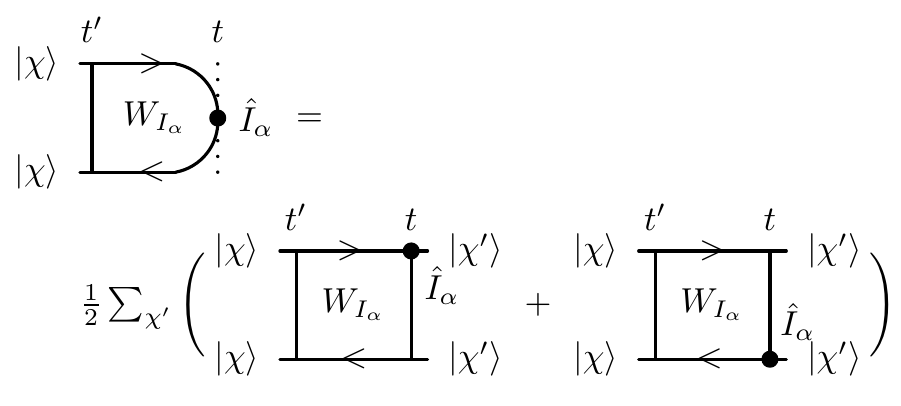}
\caption{Diagrammatic representation of the current expectation value $\langle\hat{I}_\alpha(t)\rangle$. The current operator originally placed at the turning point of the Keldysh contour, can be equivalently put on the upper or the lower contour. By summing both possibilities, a factor $1/2$ is added. \label{fig_contour_current}}
\end{figure}

In Fig.~\ref{fig_contour_current} (left-hand side), we draw a graphic representation of $W_{I_\alpha}$. The states as shown at the borders of the kernel at times $t$ or $t'$ are equal, since, as mentioned above, only diagonal elements of $P$ are of interest. The current operator sits at the turning point of the Keldysh contour at time $t$. In principle, the Keldysh contour could be continued until any later time $>t$ without changing the expectation value. Here, we choose the shortest representation. In order to directly relate $W_{I_\alpha}$ with the kernel $W$, we cut the contour at the turning point, giving the equivalent choices to place $\hat{I}_\alpha$ either on the upper or lower part of the contour (see right-hand side of Fig.~\ref{fig_contour_current}). Since both of them are equal, there is a correction factor of $1/2$ added in Eq.~(\ref{eq_def_current}). The same procedure is done for the noise auxiliary function in Eq.~(\ref{eq_noisepart_det}), hence also there appears an additional factor of $1/2$.

The resulting additional rules are set thus that all possible replacements are taken into account, i.e., the current operator has no fixed position within the diagrams. This is possible also for the current operators that are originally fixed at time $t$ (see Figs.~\ref{fig_contour_current} and \ref{fig_contour}), because we take the trace $\text{e}^\text{T}$ on all objects $W_{I_\alpha}$, $W_{I_\alpha I_\beta}$ with a fixed current operator. This trace cancels all contributions where the external vertex due to $\hat{I}_\alpha$ is not the first vertex in the diagrams, enabling us to neglect the position of $\hat{I}_\alpha$. The rules are:
\begin{enumerate}[resume]
\item Assign a factor $+1$ for each external vertex on the upper (lower) branch of the Keldysh contour which describes tunneling of an electron into (out of) lead $\alpha$, and $-1$ in the other two cases.

\item Sum up all the diagrams, taking into account the above introduced factors $\pm1$ for each possibility to replace one (two for $W_{I_\alpha I_\beta}$) tunneling vertices by current vertices. For two external vertices, multiply with a factor $1/2$.
\end{enumerate}

Finally we give the rules for the corrections of the diagrams in first order in the driving frequency $\Omega$. In the model considered here, there is no need to differ between adiabatic corrections of the kernel $W$ and of the objects $W_{I_\alpha}$ as well as $W_{I_\alpha I_\beta}$. We do in contrast separate the two cases of adiabatic corrections due to a time-dependent tunnel coupling and due to a time-depedent energy level.~\cite{Splettstoesser06}

The additional rules for adiabatic corrections $W^{\left(a\right)}$ due to the time dependence of $\Gamma$ are:
\begin{enumerate}[resume]
\item\label{rule9} Add to all diagrams needed for $W^{\left(i\right)}$ (above) additional external frequency lines between any vertex $t_{i}$ and the right corner of the diagram, and assign to them the energy $-iz_{i}$. 

\item Follow the rules (\ref{rule1}) to (\ref{rule6}) taking into account the extra lines.

\item Perform a first derivative with respect to $z_{i}$ and multiply by a factor of $\frac{1}{2}\dot{\Gamma}_\alpha\left(t\right)/\Gamma_\alpha\left(t\right)$. Sum all the contributions obtained in this way.
\end{enumerate}

The following are the additional rules for adiabatic corrections $W^{\left(a\right)}$ due to the time dependence of $\epsilon$.
\begin{enumerate}[resume]
\item In addition to the external frequency lines added according to rule (\ref{rule9}), put one more external frequency line from the left corner of the diagram with no vertex to the right corner.

\item Follow the rules (\ref{rule1}) to (\ref{rule6}) taking into account the extra lines.

\item Perform a second derivative with respect to $z_{i}$ and multiply by $-\frac{i}{2}\left(\dot{E}_{\chi}\left(t\right)\right.\left.-\dot{E}_{\chi'}\left(t\right)\right)$, where $\chi$ ($\chi'$) is the dot state entering (leaving) the vertex of the external frequency line at $t_{i}$ with respect to the Keldysh contour. The term $\dot{E}_{\chi}\left(t\right) (\dot{E}_{\chi'}\left(t\right))$ is omitted if the segment associated with $E_{\chi}\left(t\right)$ ($E_{\chi'}\left(t\right)$) does not belong to the diagram. Sum all the contributions obtained in this way.
\end{enumerate}

For the purposes of this publication, currents and current noise were calculated in the sequential tunneling regime and in the next order in an expansion in $\Gamma$. In the sequential tunneling regime, no adiabatic corrections to the kernels contribute. Therefore in the present paper no adiabatic correction rules for a time-dependent bias voltage $V$ are needed. The rules for such a general case can be found, e.g., in Ref.~\onlinecite{Kashuba12}. 
For the next-order calculation, it is furthermore enough to consider adiabatic corrections of diagrams up to first order in $\Gamma$. In this case, it is possible to formulate a simplification for the adiabatic correction in lowest order in the tunnel coupling, by relating it to the instantaneous kernel. Hence we can write down an additional simplified rule:
\begin{enumerate}[resume]
\item For the lowest-order term in $\Gamma$ simply compute $W_{t}^{\left(a,1\right)}\left(z\right)=\frac{1}{2}\partial_{z}\dot{W}_{t}^{\left(i,1\right)}\left(z\right)$.
\end{enumerate}
Note that this does not hold in general.~\cite{Kashuba12}

\section{Derivation of $\overline{\Pi}$}\label{appendix_decaying_propagator}
\subsection{Starting point}
In all noise expressions the propagator appears in an expression with a finite support, $\overline{\Pi}\left(t,t'\right)=\Pi\left(t,t'\right)-P(t)\otimes\text{e}^\text{T}$. That is, for the time difference $t-t'\rightarrow\infty$ it converges to zero. In Laplace space we can write it as
\begin{equation}
\overline{\Pi}\left(t,z\right)=\Pi\left(t,z\right)-\frac{1}{z}P\left(t\right)\otimes \text{e}^\text{T}.
\end{equation}
In order to find a relation to explicitly calculate $\overline{\Pi}$, we make a convolution with the kernel $W$ from the left-hand side, resulting in
\begin{equation}\label{eq_propLaplace}
\left(W\circ \overline{\Pi}\right)\left(t,z\right)=\left(W\circ \Pi\right)\left(t,z\right)-\left(W\circ P\right)\left(t,z\right)
\end{equation}
The convolutions in the above Eq.~(\ref{eq_propLaplace}) are evaluated according to Eq.~(\ref{eq_AB_laplace}). Since $W$ has no inverse, we need an additional condition to fully determine $\overline{\Pi}$, which is $\text{e}^\text{T}\overline{\Pi}=0$. To evaluate the convolution of the kernel $W$ and the propagator $\Pi$, we have to make use of the Dyson equation, relating propagator and kernel
\begin{equation}\label{eq_dyson_eq_time}
\Pi\left(t,t'\right)=\mathbf{1}+\int_{t'}^{t}dt_1\int_{t'}^{t_1}dt_2 W\left(t_1,t_2\right)\Pi\left(t_2,t'\right)
\end{equation}
where $W$ takes the role of a self-energy. We then transform Eq.~(\ref{eq_dyson_eq_time}) to Laplace space and obtain
\begin{equation}\label{eq_dyson_eq_laplace}
\Pi\left(t,z\right)=\frac{\mathbf{1}}{z}+\left(\mathbf{1}\circ W\circ\Pi\right)\left(t,z\right)
\end{equation}
where the last term can again be evaluated according to Eq.~(\ref{eq_AB_laplace}).
Equations~(\ref{eq_propLaplace}) and (\ref{eq_dyson_eq_laplace}) form the full set of equations we need in the following to derive the relations that determine the different contributions of $\overline{\Pi}$.

\subsection{Instantaneous contributions}
Having all ingredients in Laplace space at hand we can derive the instantaneous $\overline{\Pi}$ quite easily. We take the instantaneous limit of Eq.~(\ref{eq_propLaplace}), i.e., all time derivatives vanish and the exponentials carrying these time derivatives can be approximated by $1$. Also, the kernel $W$ as well as the density matrix $P$ are approximated to instantaneous order (i.e., zero order in the driving frequency $\Omega$). By taking in a next step the limit $z\rightarrow0^+$ we obtain
\begin{align}
W_{t}^{\left(i\right)}\overline{\Pi}_{t}^{\left(i\right)}&=W_{t}^{\left(i\right)}\Pi_{t}^{\left(i\right)}-\lim_{z\rightarrow0^{+}}\frac{1}{z}\left[W_{t}^{\left(i\right)}+z\partial W_{t}^{\left(i\right)}\right]P_{t}^{\left(i\right)}\otimes \text{e}^\text{T}\nonumber
\\\label{eq_Pbar_inst}&=
P_{t}^{\left(i\right)}\otimes \text{e}^\text{T}-\mathbf{1}-\partial W_{t}^{\left(i\right)}P_{t}^{\left(i\right)}\otimes \text{e}^\text{T}.
\end{align}
In the last line we used for one the outcome of the $z\rightarrow0^{+}$ limit of the instantaneous Dyson equation, Eq.~(\ref{eq_dyson_eq_laplace}),
\begin{equation}
W_{t}^{\left(i\right)}\Pi_{t}^{\left(i\right)}=\lim_{z\rightarrow0^{+}}z\Pi^{\left(i\right)}\left(t,z\right)-\mathbf{1}=P_{t}^{\left(i\right)}\otimes \text{e}^\text{T}-\mathbf{1}.
\end{equation}
Also, we used the result of the instantaneous master equation, $0=W_t^{(i)}P_t^{(i)}$.

The Laplace derivative of $\overline{\Pi}$ can be derived analogously by taking again the instantaneous limit of Eq.~(\ref{eq_propLaplace}). However, now we first differentiate it with respect to $z$ and then set $z$ to zero
\begin{equation}
\begin{split}\partial W_{t}^{\left(i\right)}\overline{\Pi}_{t}^{\left(i\right)}+W_{t}^{\left(i\right)}\partial\overline{\Pi}_{t}^{\left(i\right)}=\partial W_{t}^{\left(i\right)}\Pi_{t}^{\left(i\right)}+W_{t}^{\left(i\right)}\partial\Pi_{t}^{\left(i\right)}\\-\lim_{z\rightarrow0^{+}}\partial_{z}\left[\frac{1}{z}W_{t}^{\left(i\right)}+\partial W_{t}^{\left(i\right)}+\frac{1}{2}z\partial^{2}W_{t}^{\left(i\right)}\right]P_{t}^{\left(i\right)}\otimes \text{e}^\text{T}\end{split}
\end{equation}
Next, we use the relation $\overline{\Pi}_{t}^{\left(i\right)}=\partial W_{t}^{\left(i\right)}\Pi_{t}^{\left(i\right)}+W_{t}^{\left(i\right)}\partial\Pi_{t}^{\left(i\right)}$ which is a consequence of the Dyson equation, Eq.~(\ref{eq_dyson_eq_laplace}), in the instantaneous $z\rightarrow0^+$ limit. We find
\begin{equation}\label{eq_dPbar_inst}
\partial W_{t}^{\left(i\right)}\overline{\Pi}_{t}^{\left(i\right)}+W_{t}^{\left(i\right)}\partial\overline{\Pi}_{t}^{\left(i\right)}=\overline{\Pi}_{t}^{\left(i\right)}-\frac{1}{2}\partial^{2}W_{t}^{\left(i\right)}P_{t}^{\left(i\right)}\otimes \text{e}^\text{T}.
\end{equation}
Thus, Eqs.~(\ref{eq_Pbar_inst}) and~(\ref{eq_dPbar_inst}) determine the instantaneous contributions of $\overline{\Pi}$ and $\partial\overline{\Pi}$.

\subsection{Adiabatic contribution}
In this section we sketch the derivation of the adiabatic contribution of $\overline{\Pi}$. We start again from Eq.~(\ref{eq_propLaplace}), where we expand all convolutions adiabatically, according to Eq.~(\ref{eq_adiabatic_exp_general}). We then go to the $z\rightarrow0^{+}$ limit for which we get
\begin{equation}
\begin{split}\label{eq_Pi_ad_1}
\left\{ W\overline{\Pi}\right\} _{t}^{\left(a\right)}=\left\{ W\Pi\right\} _{t}^{\left(a\right)}-\lim_{z\rightarrow0^{+}}\frac{1}{z}\left\{ WP\right\} _{t}^{\left(a\right)}\otimes \text{e}^\text{T}\\-\left\{ \partial WP\right\} _{t}^{\left(a\right)}\otimes \text{e}^\text{T}.
\end{split}
\end{equation}
Likewise, we perform an adiabatic expansion of the Dyson equation for the $z\rightarrow0^{+}$ limit (note that $\lim_{z\rightarrow0^+}z\Pi^{\left(a\right)}\left(t,z\right)=P_{t}^{\left(a\right)}\otimes \text{e}^\text{T}$)
\begin{equation}\label{eq_Pi_ad_2}
P_{t}^{\left(a\right)}\otimes \text{e}^\text{T}=\left\{ W\Pi\right\} _{t}^{\left(a\right)}-\lim_{z\rightarrow0}\frac{1}{z}\dot{P}_{t}^{\left(i\right)}\otimes \text{e}^\text{T}-\dot{\overline{\Pi}}_{t}^{\left(i\right)}.
\end{equation}
Inserting Eq.~(\ref{eq_Pi_ad_2}) into Eq.~(\ref{eq_Pi_ad_1}), and using the adiabatic expansion of the generalised master equation, $\dot{P}_t^{(i)}=\left\{WP\right\}_t^{(a)}$, we arrive at
\begin{equation}\label{eq_Pbar_ad}
\left\{W\overline{\Pi}\right\}_t^{(a)}=P_t^{(a)}\otimes \text{e}^\text{T}+\dot{\overline{\Pi}}_t^{(i)}-\left\{\partial WP\right\}\otimes\text{e}^\text{T}
\end{equation}
which is the desired relation for the adiabatic correction to $\overline{\Pi}$.

\section{Analytic expressions for the current and noise including finite bias and magnetic field}\label{appendix_analytic}
In this appendix we give the explicit expressions of the prefactor matrices and vectors that appear in the analytic expressions in Sec.~\ref{sec_bias_general}.
The adiabatic current in Eq.~(\ref{eq_adiabatic_current}) contains a matrix $\mathbf{A}$. The structure of this matrix is
\begin{equation}\label{eq_A}
\mathbf{A}\left(g_{\alpha\sigma}\right)=\left(\begin{array}{cc}
\kappa\left(g_{\alpha\sigma}\right) & \zeta\left(g_{\alpha\sigma}\right)\\
\zeta\left(-g_{\alpha\sigma}\right) & \kappa\left(-g_{\alpha\sigma}\right)\end{array}\right)\ ,
\end{equation}
where the functions $\kappa$ and $\zeta$ are 
\begin{align}
\kappa\left(g_{\alpha\sigma}\right)&=\frac{1}{4Q}\sum_{\alpha,\sigma}\alpha\left(\Gamma_\alpha+\Gamma g_{\alpha\overline{\sigma}}\right)\left(g_\sigma-1\right)\label{eq_alpha}
\\
\zeta\left(g_{\alpha\sigma}\right)&=\frac{1}{2Q}\sum_{\alpha,\sigma}\alpha\sigma\Gamma_{\alpha}\left(1+g_{\alpha\sigma}\right)g_{\overline{\alpha}\overline{\sigma}}\ .\label{eq_beta}
\end{align}
The functions $\kappa$, $\zeta$ and $Q$ depend on $\Gamma_\alpha$ and $g_{\alpha\sigma}$ with the definition
\begin{equation}\label{eq_def_g}
g_{\alpha\sigma}=\frac{\Gamma_\alpha}{\Gamma}\left[f_\alpha(\epsilon_\sigma)-f_\alpha(\epsilon_\sigma+U)\right]\ .
\end{equation}
where the Fermi function for lead $\alpha$ is given as $f_\alpha\left(E\right)=1/\left(1+\exp[\beta(E-\alpha eV/2)]\right)$. 
We define $Q=\Gamma\left(1-g_{\uparrow}g_{\downarrow}\right)$ which is the product of  two new relaxation rates occurring due to the coupling of charge and spin (see, e.g. Ref.~\onlinecite{Splettstoesser10}), divided by $\Gamma$. Also we define $\bar{\alpha}=\mathrm{L}$ if $\alpha=\mathrm{R}$ and vice versa. When $\alpha$ takes the role of a variable, we have $\alpha=+1$ ($\alpha=-1$),  relating to the respective subscript $\alpha=\mathrm{L}$ ($\alpha=\mathrm{R}$) and equally for the spin, where $\sigma=+1$ ($\sigma=-1$) relates to $\uparrow$ ($\downarrow$). The off-diagonal entries $\zeta$ represent the coupling of charge and spin dynamics; therefore these coefficients have different symmetries with respect to the Zeeman splitting $\Delta$, i.e., $\zeta(-\Delta)=-\zeta(\Delta)$ whereas $\kappa(-\Delta)=\kappa(\Delta)$.

In the istantaneous noise, Eq.~(\ref{eq_instantaneous_noise}), in the term related to the current expectation value, $\vec{I}_t^{(i,1)}$, there appears the matrix
\begin{equation}\label{eq_Aprime}
\mathbf{A'}\left(g_{\alpha\sigma}\right)=-4\left(\begin{array}{cc}
\kappa\left(g_{\alpha\sigma}\right) & \zeta\left(g_{\alpha\sigma}\right)\\
\zeta\left(g_{\alpha\sigma}\right) & \kappa\left(g_{\alpha\sigma}\right)\end{array}\right)\ ,
\end{equation}
with the functions $\kappa$ and $\zeta$, as given in Eqs.~(\ref{eq_alpha}) and~(\ref{eq_beta}).
Also, there is the prefactor vector
\begin{equation}\label{eq_veca}
\vec{a}\left(g_{\alpha\sigma}\right)=\left(\begin{array}{c}
\nu\left(g_{\alpha\sigma}\right)\\
\nu\left(-g_{\alpha\sigma}\right)\end{array}\right)\ .
\end{equation}
in front of $\Delta\vec{n}_t^{(i,0)}$ in the instantaneous noise, see Eq.~(\ref{eq_instantaneous_noise}).  The function $\nu$ is found to be
\begin{equation}
\begin{split}
\nu\left(g_{\alpha\sigma}\right)=\frac{1}{Q}\left[\sum_{\alpha\sigma\sigma'}\Gamma_{\alpha}g_{\overline{\alpha}\sigma'}\frac{g_{\sigma}h_{\overline{\sigma}}}{g_{\sigma'}-g_{\overline{\sigma}'}}-\frac{\Gamma_\mathrm{L}\Gamma_\mathrm{R}}{\Gamma}h\right.\\\left.+\sum_{\sigma}g_{L\sigma}g_{R\sigma}\left(\Gamma^{2}g_{\overline{\sigma}}-\frac{2\Gamma h_{\overline{\sigma}}}{g_{\sigma}-g_{\overline{\sigma}}}\right)\right]\label{eq_gamma}
\end{split}
\end{equation}
with the additional definitions
\begin{align}
h_{\alpha\sigma}&=\Gamma_\alpha+\Gamma g_{\alpha\sigma}\\
h'_{\alpha\sigma}&=\Gamma_\alpha-\Gamma g_{\alpha\sigma}
\end{align}
Moreover we define $h_\sigma=\sum_\alpha h_{\alpha\sigma}$ and $h=\sum_\sigma h_\sigma$.
Note that for zero magnetic field, we can drop the index $\sigma$ and the auxiliary functions $h_{\alpha}$ ($h'_\alpha$) reduce to the charge (spin) relaxation rates with respect to lead $\alpha$.
Next, we list the matrices appearing in the analytic expression for the pumping noise in Eq.~(\ref{eq_sa_general}). They can be written as
\begin{equation}\label{eq_B1}
\begin{split}
\mathbf{B_1}=\frac{2}{Q}\sum_{\alpha\sigma}\left[\left(\begin{array}{cc}
0 & \sigma\Gamma_{\alpha}g_{\overline{\alpha}\overline{\sigma}}\\
0 & \Gamma g_{\alpha\sigma}g_{\overline{\alpha}\sigma}g_{\overline{\sigma}}-\frac{2\Gamma}{Q}g_{\alpha\sigma}g_{\overline{\alpha}\overline{\sigma}}h'_{\alpha\overline{\sigma}}\end{array}\right)\right.\\\left.-\frac{\Gamma_{\overline{\alpha}}h'_{\sigma}}{Q}\left(\begin{array}{cc}
0 & 0\\
0 & g_{\alpha\overline{\sigma}}+\frac{\Gamma_{\alpha}-\Gamma_{\overline{\alpha}}}{\Gamma}g_{\alpha\sigma}g_{\overline{\sigma}}\end{array}\right)\right]\mathbf{A}^{-1}
\end{split}
\end{equation}
Note that for certain cases, $\mathbf{A}^{-1}$ diverges. But because $\mathbf{B_1}$ is multiplied with $\vec{I}^{(a,0)}_t$ which itself contains $\mathbf{A}$, as can be seen in Eq.~(\ref{eq_adiabatic_current}), the term $\mathbf{B_1}\vec{I}_t^{(a,0)}$ is well-behaved. 
\begin{widetext}
The remaining matrices are
\begin{eqnarray}
\mathbf{B_2} & = & - 2\sum_{\sigma\alpha}\alpha\frac{h_{\alpha\overline{\sigma}}}{Q^{2}}
	\left(\begin{array}{cc}
		1-2g_{\sigma}+g_{\uparrow}g_{\downarrow} 
		& \sigma\left(1+g_{\sigma}\right)\\
		\sigma\left(1+2g_{\sigma}+g_{\uparrow}g_{\downarrow}\right) 
		& 1-g_{\sigma}\end{array}\right)\label{eq_B2}\\
\label{eq_B3}
\mathbf{B_3} & = & -2\sum_{\sigma\alpha}\alpha\frac{h_{\alpha\overline{\sigma}}}{Q^{2}}
	\left(\begin{array}{cc}
		1-2g_{\sigma}+g_{\uparrow}g_{\downarrow} 
		& \sigma\left(1+2g_{\sigma}+g_{\uparrow}g_{\downarrow}\right)\\
		\sigma\left(1+2g_{\sigma}+g_{\uparrow}g_{\downarrow}\right) 
		& 1-2g_{\sigma}+g_{\uparrow}g_{\downarrow}\end{array}\right)\\
\label{eq_B4}
\mathbf{B_4} & = & \sum_{\sigma\alpha}\alpha\left[\frac{\dot{\Gamma}}{\Gamma}\frac{h_{\alpha\overline{\sigma}}}{Q^{2}}
	\left(\begin{array}{cc}
		1-2g_{\sigma}+g_{\uparrow}g_{\downarrow} 
		& \sigma\left(1+2g_{\sigma}+g_{\uparrow}g_{\downarrow}\right)\\
		\sigma\left(1+2g_{\sigma}+g_{\uparrow}g_{\downarrow}\right) 
		& 1-2g_{\sigma}+g_{\uparrow}g_{\downarrow}
	\end{array}\right)\right.\\
&&\left.-\dot{g}_{\sigma}\left[h_{\alpha\overline{\sigma}}\left(1+g_{\uparrow}g_{\downarrow}\right)-2h_{\alpha\sigma}g_{\overline{\sigma}}\right]\frac{\Gamma}{Q^{3}}
	\left(\begin{array}{cc}
		-1+g_{\overline{\sigma}} 
		& \sigma\left(1+g_{\overline{\sigma}}\right)\\
		\sigma\left(1+g_{\overline{\sigma}}\right) 
		& -1+g_{\overline{\sigma}}
	\end{array}\right)\right]\nonumber
\end{eqnarray}
\end{widetext}
All that is left now are the two prefactor vectors in the pumping noise. The one which is preceeding $\Delta\dot{\vec{n}}_t^{(i,0)}$ can
be given as
\begin{align}\label{eq_b1}
\vec{b}_{1}=&\frac{\Gamma_\mathrm{L}\Gamma_\mathrm{R}}{\Gamma^{2}}\left[\frac{1}{2Q}\left(\begin{array}{c}
h'\\
-h\end{array}\right)+\frac{\Gamma}{Q^{2}}\left(\begin{array}{c}
4\Gamma g\\
2h'-Qg\end{array}\right)\right]\\\nonumber&+\sum_{\alpha\sigma}\Gamma_{\alpha}g_{\overline{\alpha}\sigma}\left[\frac{2\Gamma}{Q^{2}}\left(\begin{array}{c}
g\\
1\end{array}\right)+\sigma\frac{2}{\Gamma C}\left(\begin{array}{c}
-hg_{\sigma}\\
2g_{\overline{\sigma}}h'_{\sigma}\end{array}\right)\right]\\\nonumber&-\sum_{\sigma}\left\{ g_{L\sigma}g_{R\overline{\sigma}}\frac{\Gamma}{2Q^{2}}\left(\begin{array}{c}
4\Gamma+Q\\
Q\end{array}\right)\right.\\\nonumber&+\left.g_{L\sigma}g_{R\sigma}\left[\frac{\Gamma g_{\overline{\sigma}}}{2Q^{2}}\left(\begin{array}{c}
Q\\
3Q+4\Gamma\end{array}\right)+\sigma\frac{4}{C}\left(\begin{array}{c}
h_{\overline{\sigma}}\\
-h'_{\overline{\sigma}}\end{array}\right)\right]\right\} 
\end{align}
The second vector, $\vec{b}_2$, is too big to be displayed in this paper. It is of similar shape as $\vec{b}_1$, but contains terms that are proportional to the time derivatives $\dot{\Gamma}_\alpha$ and $\dot{g}_{\alpha\sigma}$. We want to stress however, that all results were obtained analytically, hence there are no additional approximations apart from the adiabatic expansion made in Sec.~\ref{sec_adiabatic} and the lowest-order expansion in the tunnel coupling in Sec.~\ref{sec_tunnel_exp}.

\end{document}